\documentclass[aps,prl,twocolumn,longbibliography,superscriptaddress]{revtex4-2}
\usepackage{amssymb}
\usepackage{amsbsy}
\usepackage{amsmath}
\usepackage{graphicx}
\usepackage{graphics}
\usepackage{setspace}
\usepackage{array}
\usepackage{color}
\usepackage{fontenc}
\usepackage{textcomp}
\usepackage{bm}
\usepackage{float}
\usepackage{pifont}
\usepackage[bookmarks=false,linkcolor=blue,urlcolor=blue,colorlinks,citecolor=blue]{hyperref}
\usepackage{soul}
\usepackage[T1]{fontenc}
\usepackage[percent]{overpic}

\begin{document}
\author{T. Farajollahpour}
\email{tohid.f@brocku.ca}
\affiliation{Department of Physics, Brock University, St. Catharines, Ontario L2S 3A1, Canada}

\author{R. Ganesh}
\email{r.ganesh@brocku.ca}
\affiliation{Department of Physics, Brock University, St. Catharines, Ontario L2S 3A1, Canada}

\author{K. V. Samokhin}
\email{ksamokhin@brocku.ca}
\affiliation{Department of Physics, Brock University, St. Catharines, Ontario L2S 3A1, Canada}

\date{\today}
	
\title{Berry curvature-induced transport signature for altermagnetic order} 

\begin{abstract}
  Altermagnetism has been detected in several materials using spin-sensitive probes. These measurements require rather complex setups that make it challenging to track variations in altermagnetic order, e.g., to identify a temperature-tuned altermagnetic phase transition. We propose a simple transport measurement that can probe the order parameter for $d$-wave altermagnetism. We suggest magnetoconductivity anisotropy -- the difference between the two principal values of the magnetoconductivity tensor. This quantity can be easily measured as a function of temperature, without any spin-selective apparatus. It acquires a nonzero value in a $C_4K$ phase, where $C_4$ rotations and time reversal $K$ are not symmetries but their combination is. 
This effect can be traced to the modification of phase space density due to Berry curvature, which we demonstrate using semiclassical equations of motion for band electrons. As an illustration, we build a minimal tight-binding model with altermagnetic order that breaks $C_4$ and $K$ symmetries while preserving $C_4K$. 
\end{abstract}

\maketitle

\maketitle 
Transport measurements have emerged as powerful probes of band topology in condensed matter systems~\cite{BerryOriginal,Berry1app,Berry2app,Berry3app,Berry4app,Berry5app,Berry6app,BerryReview}. This is due to the strong effect of the Berry curvature on charge carrier dynamics, analogous to external electromagnetic fields~\cite{Sundaram,BerryReview}. 
In the semiclassical transport regime, the Berry curvature effects arise from two factors: (i) an anomalous contribution to wavepacket velocity and (ii) a correction to the density of states in phase space~\cite{DensityBliokh,momentBliokh2006,DensityDuval2006,NiuPRL2007}.
The former has received considerable attention, e.g., as providing an intrinsic contribution to various transport properties~\cite{Berry5app,QHEreview2023}. The latter is relatively less studied, especially in the context of electronic transport. 
Here, we demonstrate that this latter effect leads to a magnetoconductance anisotropy in $C_4K$ materials.  We interpret this quantity as an order parameter which can be used to identify an altermagnetic phase transition. 

It is well known that the Berry curvature is highly constrained by symmetries. A Berry curvature monopole can only appear in systems that break time reversal (TR) symmetry. A Berry dipole requires breaking of inversion symmetry~\cite{LiangFu2015}. Recent studies have shown that a Berry curvature quadrupole can appear in systems with $C_4K$ symmetry~\cite{LawPRB2023,Tohid2024}. Here, both TR symmetry ($K$) and a fourfold rotational symmetry ($C_{4}$) are broken, while their combination $C_{4}K$ remains a symmetry. In such systems, there is no Berry curvature-induced conductivity in linear or quadratic orders in the applied electric field. It appears only at the third order in the applied electric field. 
In this paper, we focus on magnetoconductivity anisotropy -- a transport signature that requires both external electric and magnetic fields.  
Crucially, being related to the Berry quadrupole, this quantity is present in $d$-wave altermagnetic materials that exhibit $C_{4}K$ symmetry~\cite{SinovaPRX2022,Sinova2PRX2022,Fedchenko_undated-qm}. 
Intriguingly, the very same symmetry requirements are also invoked in chiral 
higher-order topological crystalline insulators~\cite{Schindler2018}. Here, we restrict our attention to metallic systems where conductivity can be viewed as a Fermi-surface property~\cite{Berry6app,Wang2007}. 

\textbf{Results}

\begin{figure}
    \centering
    \includegraphics[width=0.7\columnwidth]{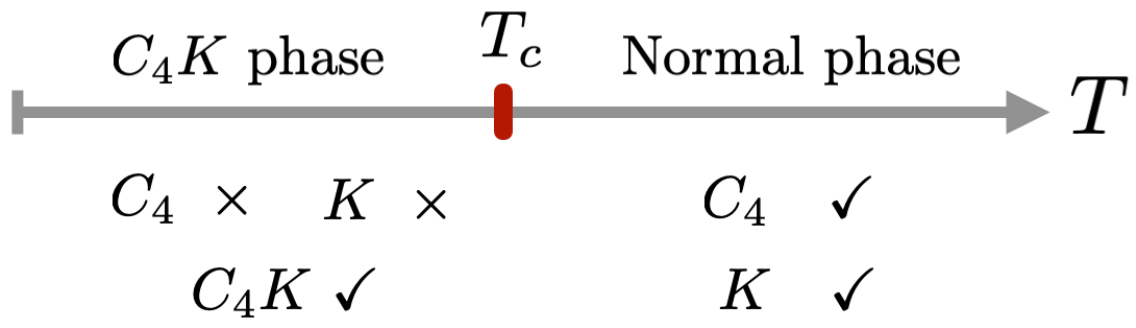}
\caption{Symmetry change across a temperature-tuned $d$-wave altermagnetic phase transition. }  
\label{fig:Setup}
\end{figure}

We consider a transport setup where the sample is a metal with $C_4K$ symmetry. We apply a weak external magnetic field $\bm{B}=B_z\hat{\bm{z}}$, with the $z$ axis taken to coincide with the fourfold axis. We apply an external dc electric field $\bm E$ in the $xy$ plane. From current measurements, we extract the conductivity tensor $\hat\sigma(B_z)$ as follows:
\begin{equation}
    \left(\begin{array}{c} j_x \\ j_y \end{array}
    \right) = 
    \left(\begin{array}{cc}
    \sigma_{xx} & \sigma_{xy} \\
    \sigma_{yx} & \sigma_{yy} 
    \end{array}\right)\left(\begin{array}{c} E_x \\ E_y \end{array}
    \right) + \mathcal{O}(E^2).
\label{eq:linear-resp}
\end{equation}
This tensor is non-diagonal, in general, so we seek to calculate its principal values. This can be done by rotating the measurement apparatus about the $z$ axis ($x, y \rightarrow x', y'$), until the off-diagonal elements are equal and opposite in sign, i.e., $\sigma_{x'y'} = -\sigma_{y'x'}$. This corresponds to a transverse response that has a pure Hall character. The longitudinal response is captured by the diagonal entries, $\sigma_{x'x'}$ and $\sigma_{y'y'}$, which constitute the principal values. We focus on the anisotropy parameter, defined as 
\begin{eqnarray}
\eta = \vert \sigma_{x'x'}(B_z)-\sigma_{y'y'}(B_z)\vert.
\label{Eq.eta}
\end{eqnarray}
In practice, the anisotropy can be determined without rotating the apparatus -- by applying the electric field along two perpendicular axes, $x$ and $y$, and measuring currents along $x$ and $y$ in each case -- see discussion below. 

Below, we focus on the weak field limit and expand the conductivity tensor as follows:
\begin{equation}
\sigma_{ij}(B_z) = \tilde\sigma_{ij} + \alpha_{ij} B_z + \mathcal{O}(B_z^2),
\end{equation}
where the first term represents the conductivity in the absence of a magnetic field. We refer to the magnetic field-dependent contribution, $\alpha_{ij}$, as the magnetoconductivity tensor. 

We separate each element of $\tilde\sigma_{ij}$ and $\alpha_{ij}$ into two contributions: one that does \textit{not} involve Berry curvature and one that does. The former ones, designated as ``Drude'' ($\text{D}$), are given by
\begin{eqnarray}
\tilde\sigma_{ij}^{\text{D}} &=& -\tau \frac{e^2}{\hbar} \int \frac{d^d \bm k}{(2\pi)^d}\, v_i v_j \frac{\partial f_0}{\partial \varepsilon},
\label{Eq.Drude1}\\
\alpha_{ij}^{\text{D}} &=& -\tau^2 \frac{e^3}{\hbar}  \int \frac{d^d \bm k}{(2\pi)^d}\, \epsilon_{\ell n 3} v_i v_n \frac{\partial v_j}{\partial k_\ell}  \frac{\partial f_0}{\partial \varepsilon}.~~~
\label{Eq.Drude}
\end{eqnarray}
Here $\alpha^{\text{D}}_{xy}$ represents the conventional Hall conductivity, which originates from Lorentz force \cite{hurd2012hall,Ziman1979}, and $\tau$ is the relaxation time. 
The Berry curvature-dependent contributions ($\text{B}$) have the form
\begin{eqnarray}
    && \tilde\sigma_{xx}^{\text{B}} =\tilde\sigma_{yy}^{\text{B}} = 0, \\
    && \tilde\sigma_{xy}^{\text{B}} = -\tilde\sigma_{yx}^{\text{B}}  = -\frac{e^2}{\hbar} \int \frac{d^d \bm k}{(2\pi)^d}\,\Omega_z f_0,
\end{eqnarray}
where $\Omega_z$ is the Berry curvature, see Methods section below. The second line here corresponds to the anomalous Hall conductivity~\cite{Nagaosa2010,QHEreview2023}. For the Berry-curvature contribution to the magnetoconductivity we obtain
\begin{equation}
    \alpha_{ij}^{\text{B}} = \tau \frac{e^3}{\hbar} \int \frac{d^d \bm k}{(2\pi)^d}\, v_i  v_j  \Omega_z \frac{\partial f_0}{\partial \varepsilon}.~~~
\label{Eq.Berry}
\end{equation}
Here, in the interest of brevity, we have suppressed contributions  that are proportional to the magnetic moment of the wavepacket. In the Supplementary Material~\cite{supplementary}, we provide explicit expressions and argue that they have the same symmetry properties as the contribution from Eq.~(\ref{Eq.Berry}). 
Notably, $\alpha_{ij}^{\text{B}}$ above originates from the phase-space-volume modification factor $D(\bm k)$ (see Methods). From the expression (\ref{Eq.Berry}), we see that this quantity is related to the quadrupole moment of the Berry curvature.

\begin{table}[]
    \centering
    \begin{tabular}{c||c|c|c|c|c}
       &$v_x$ & $v_y$  & $\Omega_z$ & $\bm \int_{\bm k} \Omega_z$ & $\bm \int_{\bm k} v_xv_y\Omega_z$\\ 
       \hline \hline$K$ & $-v_x$ & $-v_y$  & $-\Omega_z$ &  odd & odd\\
      \hline   $C_4$& $v_y$ & $-v_x$ & $\Omega_z$&  even & odd\\
    \hline   $C_2$& $-v_x$ & $-v_y$ & $\Omega_z$&  even & even\\
      \hline   $C_4 K$& $-v_y$ & $v_x$& $-\Omega_z$&  odd & even\\
      \hline
    \end{tabular}
    \caption{Symmetry properties of various physical quantities and their integrals under 
    different symmetry operations. The terms ``even'' and ``odd'' indicate the behavior of 
    the integrands under the respective transformations.}
    \label{table1}
\end{table}

\textbf{Symmetry arguments}

We consider a material that undergoes a temperature-tuned altermagnetic phase transition as shown in Fig.~\ref{fig:Setup}. In the ``normal'' phase above $T_c$, the material has both $C_4$ and $K$ symmetries. In the altermagnetic phase below $T_c$, $C_4$ and $K$ are broken, but $C_4K$ is preserved. In addition, the material has $C_2$ symmetry, as applying $C_4K$ twice results in $C_2$. The symmetry properties of the velocity $\bm{v}$ and the Berry curvature ${\Omega}_z$ are summarized in Table~\ref{table1}. 

In the normal phase, $C_4$ symmetry immediately forces $\eta=0$, i.e., there can be no conductivity anisotropy. 
In the altermagnetic phase, we consider the expressions obtained from the Boltzmann transport in Eqs.~(\ref{Eq.Drude1}-\ref{Eq.Berry}). These expressions are strongly constrained by $C_4K$ symmetry. Namely, for the Drude contributions we have
\begin{equation}
\label{eq:cond-Drude}
    \begin{array}{l}
    \tilde\sigma_{xx}^{\text{D}} =  \tilde\sigma_{yy}^{\text{D}},\quad
  \tilde\sigma_{xy}^{\text{D}} =  \tilde\sigma_{yx}^{\text{D}} =0,\medskip\\
  \alpha_{xx}^{\text{D}} =  \alpha_{yy}^{\text{D}},\quad
  \alpha_{xy}^{\text{D}} =  -\alpha_{yx}^{\text{D}}.
    \end{array}
\end{equation}
For the Berry curvature-induced contributions to magnetoconductivity, we obtain 
\begin{equation}
\label{eq:magn-cond-Berry}
    \alpha_{xx}^{\text{B}} = -\alpha_{yy}^{\text{B}},\quad
    \alpha_{xy}^{\text{B}} = \alpha_{yx}^{\text{B}}.
\end{equation}
Note that the Berry curvature-induced contribution is purely symmetric. 

To extract the anisotropy in conductivity, we express the conductivity tensor in Eq. (\ref{eq:linear-resp}) as
\begin{eqnarray}
\hat \sigma = \sigma_0 \hat \tau_0 + \sigma_1 \hat \tau_1 + \sigma_2 \hat \tau_2 + \sigma_3 \hat \tau_3,  
\label{EQ.Paulis}
\end{eqnarray}
where $\hat\tau_0$ is the $2 \times 2$ identity matrix and $\hat\tau_{1,2,3}$ are the Pauli matrices. Here,
\begin{itemize}
    \item $\sigma_0$ represents the rotationally symmetric longitudinal conductivity arising from $\hat{\tilde\sigma}^{\text{D}}$;
    \item $\sigma_1$ is the symmetric transverse response arising from $\hat\alpha^{\text{B}}$;
    \item $\sigma_2$ represents the anti-symmetric transverse conductivity or the Hall response, originating from $\hat\alpha^{\text{D}}$ (note that $\sigma_2$ is purely imaginary);
    \item $\sigma_3$ is the anisotropic longitudinal response, also arising from $\hat\alpha^{\text{B}}$.
\end{itemize}
To identify the principal axes of the conductivity tensor, we rotate the setup about the $z$ axis by an angle $\theta$. This is achieved by a transformation of the form $\hat{\sigma}'=e^{-i\theta\hat\tau_2}\hat{\sigma}e^{i\theta\hat\tau_2}$. With an appropriate choice of $\theta$, we obtain
\begin{eqnarray}
    \hat\sigma' = \sigma_0 \hat\tau_0 +  \sigma_2 \hat\tau_2 + \sqrt{\sigma_1^2+\sigma_3^2}\,\hat\tau_3. 
\label{EQ.Paulisprinc}
\end{eqnarray}
The off-diagonal part, given by $\sigma_2\hat\tau_2$, is purely anti-symmetric. It represents a Hall response that cannot be removed by rotating the axes. The $\sigma_0\hat\tau_0$ term represents an isotropic response. 

We identify $\sqrt{\sigma_1^2+\sigma_3^2}$ as the source of the conductivity anisotropy, defined in Eq.~(\ref{Eq.eta}). 
Remarkably, according to the $C_4K$ symmetry constraints, see Eqs. (\ref{eq:cond-Drude}) and (\ref{eq:magn-cond-Berry}), it arises solely due to the Berry curvature-induced component of magnetoconductivity:
\begin{equation}
\label{eq:eta-alpha-B}
    \eta = 2\vert B_z \vert \sqrt{\left(\alpha_{xy}^{\text{B}}\right)^2 + \left(\alpha_{xx}^{\text{B}}\right)^2}.
\end{equation}
We note that, in order to determine $\eta$, it suffices to apply an electric field along two directions: $x$ and $y$. This allows for measuring all four components of the conductivity tensor. By recasting it in the form of Eq.~(\ref{EQ.Paulis}), one can immediately find $\eta$.

In summary, $\eta$ vanishes in the normal phase above $T_c$. It acquires a non-zero value in the ordered altermagnetic phase -- due to the Berry curvature-induced magnetoconductivity of Eq.~(\ref{Eq.Berry}). Crucially, $\eta$ can be measured with a simple transport apparatus, without any spin-selective components.  Thus, $\eta$ can serve as an easily-measurable ``order parameter'' for the altermagnetic transition.

\begin{figure}
    \centering
 \includegraphics[width=1\columnwidth]{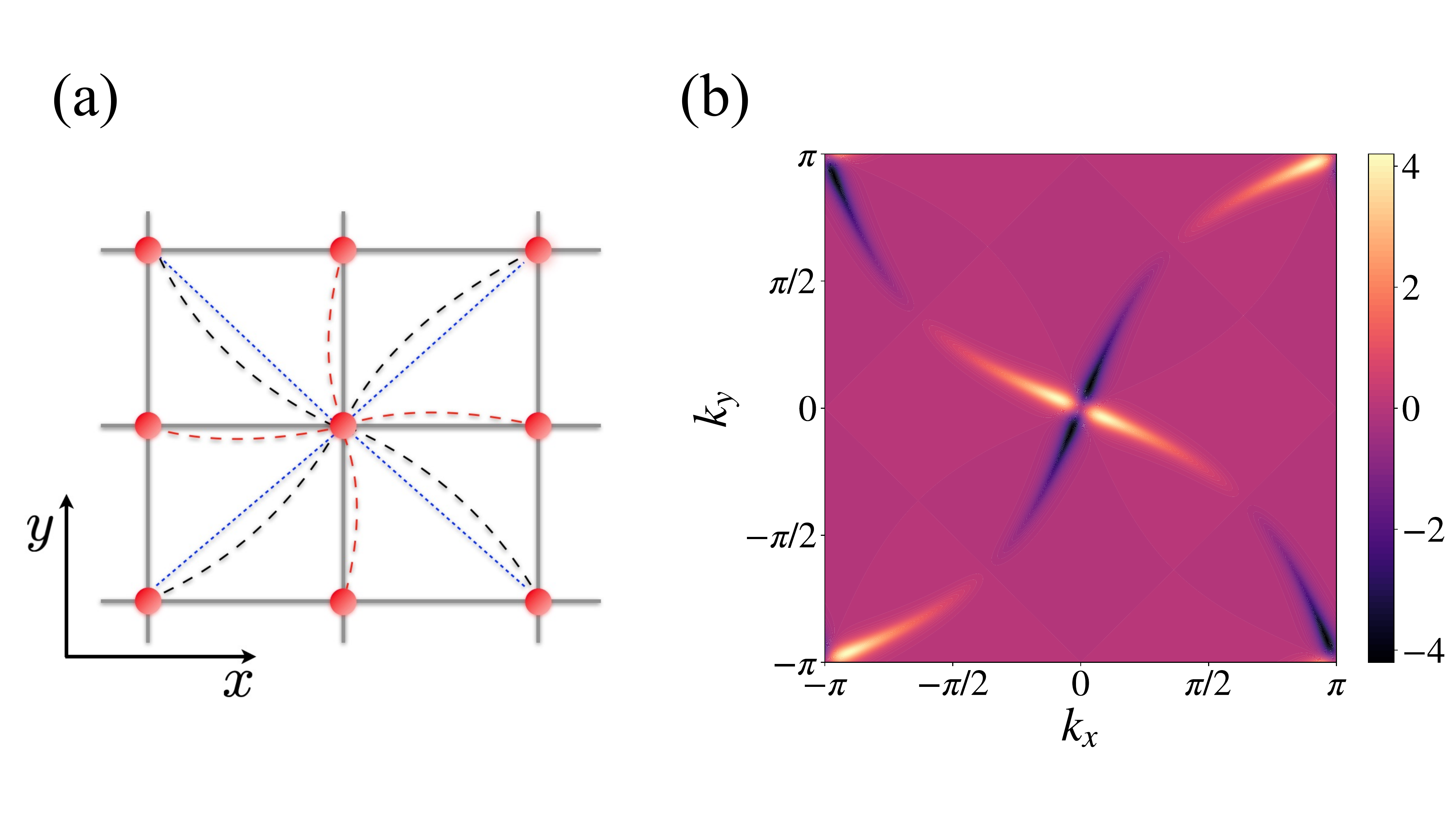}
\caption{(a) Proposed tight-binding model for a $C_4 K$ material. On an underlying square lattice, we have standard hopping processes between nearest neighbours. Along diagonals, 
we have Rashba-like hoppings shown as blue dotted lines. Altermagnetic order is captured in two spin-dependent hopping processes: $J_1$ between nearest neighbours (red dashed lines) and $J_2$ between next nearest neighbours (black dashed lines). 
(b) The $C_4K$ symmetry of the altermagnetic phase can be seen in the Berry curvature distribution, shown here for the upper band. 
}
\label{Tight-Berry}
\end{figure}

\textbf{Toy model}

As a minimal model for a $C_{4}K$ material, we consider a two-band system in two dimensions (2D) described by a Hamiltonian $\hat H(\bm k)=d_0({\bm k})\tau_0+ {\bm d}(\bm k)\cdot\hat{\bm\tau}$, where $\hat{\bm\tau}$ is a vector of the Pauli matrices that act on the spin degree of freedom.
We have two bands with energies given by $d_0({\bm k}) \pm \vert \bm{d}({\bm k})\vert$.
At momenta that are invariant under $C_4K$, we must necessarily have ${\bm d}(\bm k)=\bm{0}$, corresponding to a band degeneracy. In the neighbourhood of each such point, we have the following effective model~\cite{Tohid2024}:
\begin{eqnarray}
   d_0(\bm k) &=& a_0 + a_1( k^2_x + k^2_y ),\nonumber\\
     d_1(\bm k) &=& b_1 k_x+b_2 k_y, \nonumber\\
     d_2(\bm k) &=&-b_2 k_x + b_1 k_y , \nonumber\\ 
     d_3(\bm k) &=& m_1 (k_x^2-k_y^2) + 2m_2 k_xk_y,
    \label{Eq:ConstraintsSpin}
\end{eqnarray}
where $\bm{k}$ is measured from the degeneracy point. This is the most general form of a two-band Hamiltonian that is consistent with the antiunitary $C_{4}K$ symmetry. If additional symmetries are present, then some of the terms vanish. For example, if the symmetry group below $T_c$ contains an additional mirror reflection $\sigma_y$, then $b_1=m_1=0$.

To better understand the origin of various terms in the Hamiltonian, we construct a tight-binding model that reproduces Eq.~(\ref{Eq:ConstraintsSpin}) near the degeneracy points. As illustrated in Fig.~\ref{Tight-Berry}, apart from the usual hopping processes, we have spin-dependent hoppings between next-nearest neighbours which arise from the Rashba spin-orbit coupling. We also introduce ``altermagnetic order parameters'', $J_1$ and $J_2$, which encode preferential hopping of each spin along nearest and next-nearest neighbour bonds. Crucially, the $J_1$ and $J_2$ processes break both $C_4$ and $K$ symmetries, but preserve $C_4 K$. The Hamiltonian has the following form in momentum space:   
\begin{eqnarray}
    && \hat H(\bm k) = -t \big(\cos k_x   + \cos  k_y  \big) \hat\tau_0 \nonumber\\ 
    && \qquad +\frac{\lambda}{2} \left[\sin  \left(k_x +k_y \right)\hat\tau_1 + \sin \left(k_y-k_x \right) \hat\tau_2 \right] \nonumber\\ 
  && \qquad +\left[ J_1(\cos  k_x  - \cos  k_y ) + J_2 \sin  k_x  \sin k_y \right]\hat\tau_3,
\label{Eq:HamiltonianMomentum}
\end{eqnarray}
where $t$ denotes the hopping parameter and $\lambda$ corresponds to the Rashba spin-orbit coupling. We set the lattice constant to unity. The band spectrum has Dirac points at two $C_4K$-invariant momenta: $\bm k=(0,0)$ and $(\pi,\pi)$, corresponding to $\Gamma$ and $M$ points in the Brillouin zone respectively. 

\begin{figure}[t]
    \centering
    \includegraphics[width=1\columnwidth]{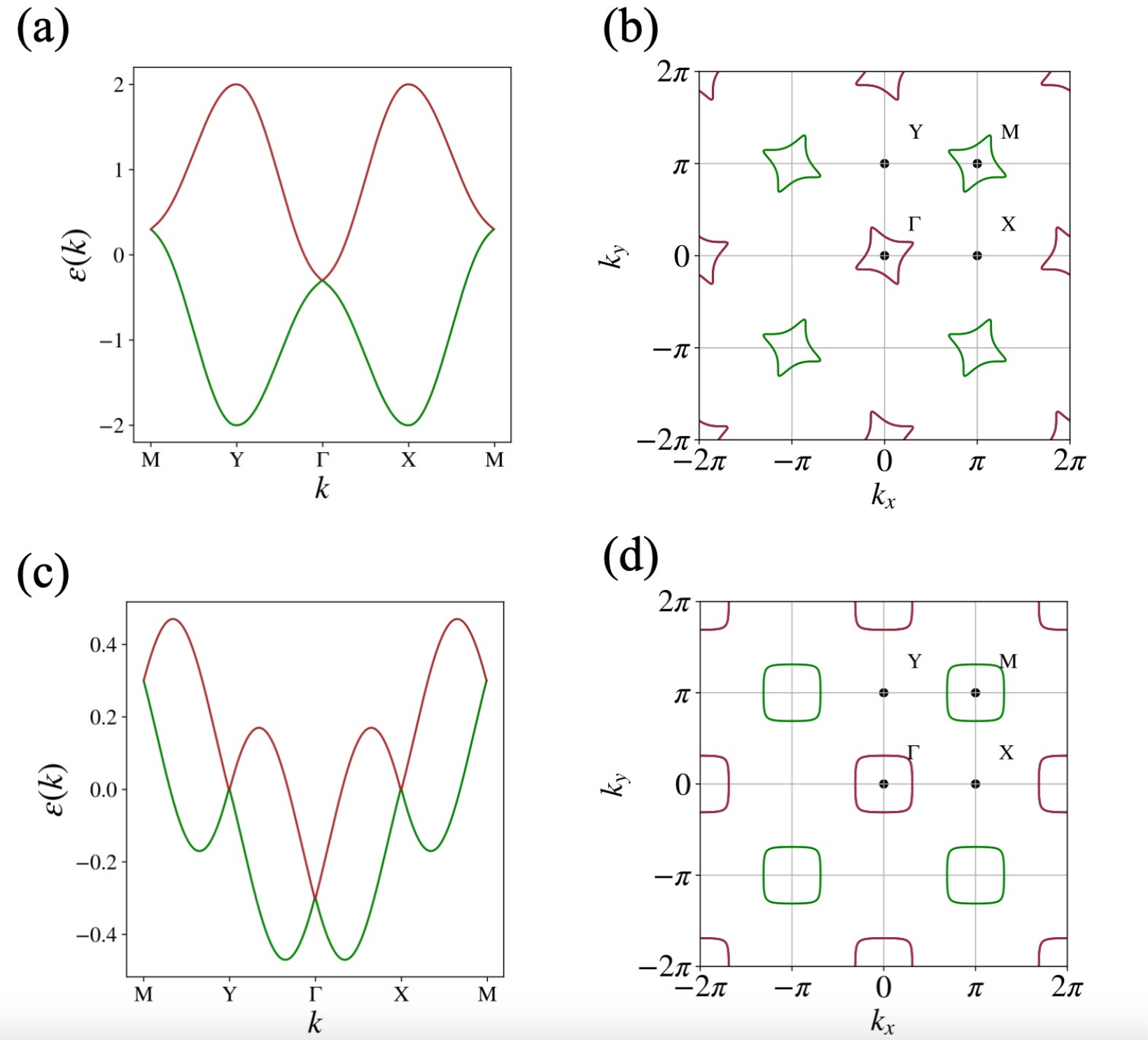}
\caption{(a) Band structure along a high-symmetry contour and (b) Fermi surfaces in the altermagnetic phase with $J_1=J_2\neq 0$. (c, d) Band structure and Fermi surfaces without altermagnetic order, i.e., with $J_1=J_2=0$. The chemical potential $\mu$ is set to zero. 
}    
\label{fig:bandFermi}
\end{figure}

Energy bands along a high-symmetry path in the Brillouin zone are shown in Fig.~\ref{fig:bandFermi}(a,c). 
The $\Gamma$ and $M$ points always host gapless Dirac nodes. Without altermagnetic order ($J_1=J_2=0$), the system is invariant under both $C_4$ and $K$, and two additional gapless Dirac nodes appear at $X$ and $Y$. The band degeneracies at $X$ and $Y$ are removed by altermagnetic order. At zero chemical potential, two Fermi pockets form around each Dirac point, as shown in Fig.~\ref{fig:bandFermi}(b,d). 

\textbf{Conductivity tensor in the toy model}

We now derive analytic expressions for conductivities. Starting from the tight-binding bands obtained from Eq.~(\ref{Eq:HamiltonianMomentum}) and focusing on the vicinity of each Dirac point, we recover the long-wavelength form of Eq.~(\ref{Eq:ConstraintsSpin}) -- see the Supplementary Material~\cite{supplementary}. Around the $\Gamma$ point, the model parameters are given by 
$a_0 =-2t $, $a_1=t/2$, $b_1 = b_2 =\lambda /2$, $m_1 =-J_1/2$, $m_2 =J_2/2$. Near the $M$ point, we have $a_0 =2t$, $a_1=t/2$, $b_1 = b_2 = \lambda/2$, $m_1 =J_1/2$, and $m_2=J_2/2$. 

At each Dirac node, we have an upper band and a lower band. Their Berry curvatures are given by
\begin{align}
 &  \Omega^{\zeta}_{z,\pm} =  \pm \zeta \frac{ [J_1 (k_x^2 - k_y^2) -2 \zeta J_2  k_x k_y] \lambda^2  }{\left( [J_1  (k_x^2 - k_y^2)-2\zeta J_2  k_xk_y]^2 +2 \lambda^2  k^2 \right)^{3/2}},
\end{align}
where $\zeta =+1$ around $\Gamma$ and $-1$ around $M$. The upper (lower) sign applies for the upper (lower) band. Note that the Berry curvature is nonzero only if \textit{both} altermagnetic order and the spin-orbit coupling are present. 

We assume that the Fermi energy is close to both Dirac points, resulting in two small Fermi pockets. For concreteness, we suppose that the Fermi energy crosses only the lower band at each pocket. We further assume weak altermagnetic order with  $J_1=J_2 = J\ll \lambda$, which results in nearly circular Fermi surfaces (we find the same qualitative behaviour when $J_1 \neq J_2$ \cite{supplementary}). Finally, assuming that temperature is much smaller than other energy scales, such as the bandwidth and the Fermi energy, we derive analytic expressions for the components of the conductivity tensor to leading order in $J/\lambda$. 

\begin{figure}
\centering
    \includegraphics[width=.8\columnwidth]{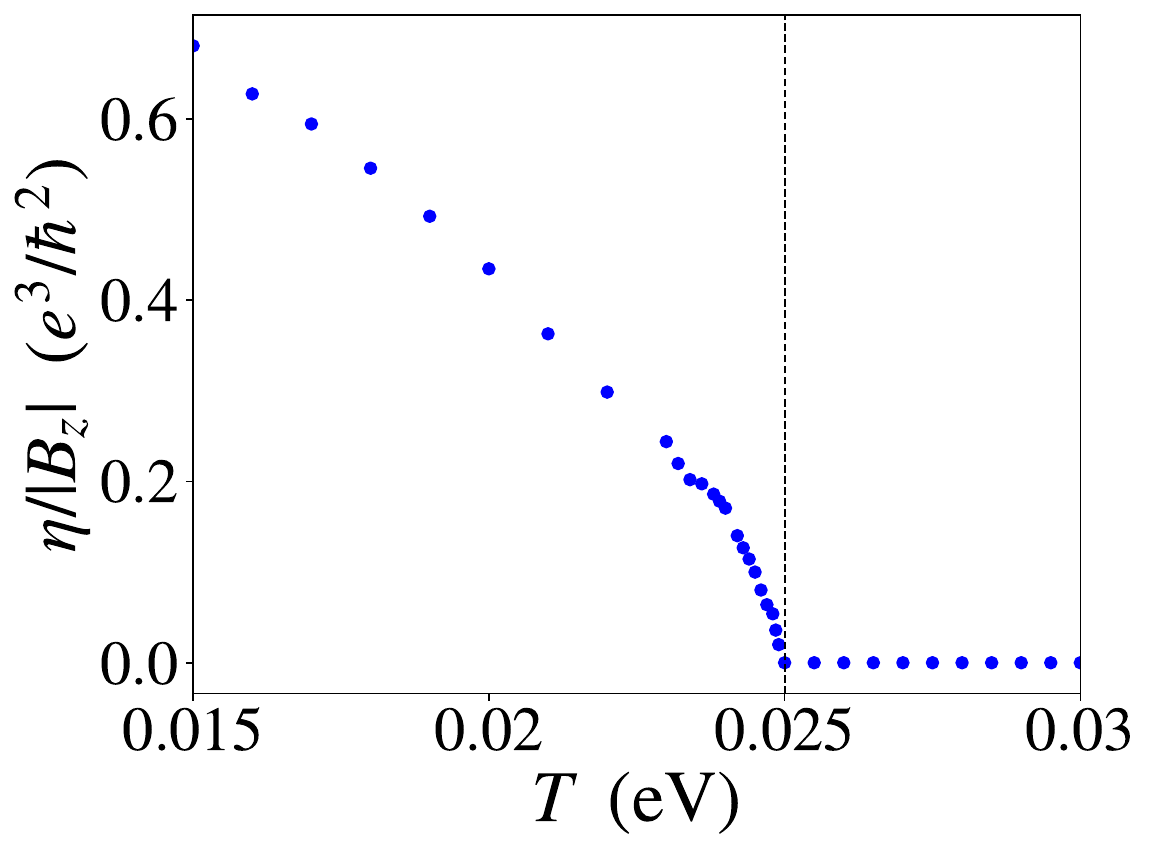}
    \caption{Evolution of the magnetoconductivity anisotropy across an altermagnetic phase transition. The anisotropy is calculated numerically from the tight-binding model given by Eq.~(\ref{Eq:HamiltonianMomentum}). The chemical potential is set at $\mu=0$. The other parameters used are $\lambda = 0.4~\text{eV}$, $t = 0.15~\text{eV}$, $J_0 = 1~\text{eV}$, and $\hbar/\tau = 0.066~\text{eV}$. The dashed line indicates the critical temperature $T_c$.}
    \label{fig:Sigma}
\end{figure}

Focusing on the anisotropy arising from the Berry curvature-induced mangnetoconductivity, we obtain
\begin{equation}
\label{eq:model-alpha-B}
    \alpha_{xx}^{\text{B}} =- \alpha_{yy}^{\text{B}} = 2\alpha_{xy}^{\text{B}} = 2 \alpha_{yx}^{\text{B}} = CJ,
\end{equation}
where $J$ is the altermagnetic order parameter and 
\begin{equation}
    C =\frac{\tau e^3 }{16\pi \hbar \lambda^2}
 \left[
\frac{(\lambda^2-2t\epsilon_\Gamma)^2}{\lambda^2+2t\epsilon_\Gamma} -\frac{(\lambda^2-2t\epsilon_M)^2}{\lambda^2+2t\epsilon_M }
 \right],
 \label{eq:C}
\end{equation}
with $\epsilon_\Gamma= \vert 2t+\mu\vert$ and $\epsilon_M= \vert 2t-\mu\vert$ being the energies of degenerate bands at the Dirac points. It is easy to see that the above expressions for the magnetoconductivity satisfy the Onsager relations \cite{LL-5}. Indeed, since both the applied magnetic field and the altermagnetic order parameter are odd under TR, one must have $\sigma_{ij}(B_z,J)=\sigma_{ji}(-B_z,-J)$, i.e., $\alpha_{ij}(J)=-\alpha_{ji}(-J)$. 

The expressions (\ref{eq:model-alpha-B}) contribute to the $\sigma_1$ and $\sigma_3$ coefficients as defined in Eq.~(\ref{EQ.Paulis}). From Eq. (\ref{eq:eta-alpha-B}), we obtain the magnetoconductivity anisotropy as
\begin{eqnarray}
\eta = \sqrt{5}~\vert C B_z\vert \, \vert J\vert.
    \label{eq.toy_eta}
\end{eqnarray}
We see that $\eta$ is directly proportional to the altermagnetic order parameter. As we cross an altermagnetic phase transition from the normal phase to the ordered phase, $\eta$ will develop a non-zero value.

\textbf{Robustness of the Berry curvature contribution}

To derive Eq.~(\ref{eq.toy_eta}), we have used the long-wavelength model given by Eq.~(\ref{Eq:ConstraintsSpin}) with several assumptions. This provides an analytic expression for the conductivity anisotropy when the Fermi energy is close to the Dirac points. Next, we calculate the conductivities numerically, relaxing all assumptions. We work directly with the tight-binding model of Eq.~(\ref{Eq:HamiltonianMomentum}). To study dependence on temperature, we suppose that the altermagnetic order sets in via a second-order phase transition at a critical temperature $T_c$. Following standard Landau theory, we assume that the altermagnetic order parameter is given by \cite{LL-5}
\begin{align}
J(T) = J_0 \sqrt{1 - \frac{T}{T_c}} \quad \text{for } T < T_c,
\label{eq.JT}
\end{align}
where $J_0$ is a constant that depends on the material parameters. As $T$ approaches $T_c$ from below, the system transitions to the normal state (with $C_4$ and $K$ symmetries) where altermagnetic order vanishes. 

The results of a numerical calculation of $\eta$ as a function of temperature are plotted in Fig.~\ref{fig:Sigma}. 
At the critical temperature, $\eta$ shows a pronounced square-root singularity with a divergent slope.
This is a robust feature, regardless of details such as the strength of altermagnetic order, chemical potential, etc.

\textbf{Discussion}

Altermagnets are usually defined as materials that retain spin polarization even in the absence of spin-orbit coupling. Nevertheless, any candidate material will have some spin-orbit coupling, however weak. For example, RuO$_2$ has been estimated to have a spin-orbit coupling of $\sim$160 meV~\cite{Occhialini2021}. This will inevitably lead to a nonzero Berry curvature, which will, in turn, give rise to a magnetoconductivity anisotropy with a singular temperature dependence. We have presented a two-dimensional toy model which allows for analytic calculation of this effect. Our arguments also apply to three-dimensional materials, as they only rely on $C_4K$ symmetry (with the magnetic field aligned along the symmetry axis). 

Our results can be readily tested in 
materials such as $\rm{KV_2Se_2O}$~\cite{Jiang2025}, $\rm{RuO_2}$, $\rm{MnO_2}$, and $\rm{MnF_2}$~\cite{Sinova2PRX2022,SinovaPRX2022}.
Our proposal is based on $C_4K$ symmetry, which corresponds to $d$-wave altermagnetism. An exciting future direction is to examine whether similar transport signatures exist in other altermagnetic systems, e.g., in $g$-wave altermagnets. 

\textbf{Methods}

We consider the semiclassical description of transport in the presence of external electric and magnetic fields, $\bm E$ and $\bm B$. 
The steady-state electric current density in a spatially uniform system can be expressed in 
terms of the distribution function $f({\bm k})$ as 
\begin{align}
     j_i = -e \int \frac{d^d{\bm k}}{(2\pi)^d}\, D(\bm k)\, {\dot{r}}_i\, f(\bm k), 
     \label{Eq:current}
\end{align}
where $d$ is the dimensionality of the system and the electron charge is $-e$. The measure $D(\bm k)$, which encodes a modification of the phase space volume arising from the noncanonical dynamics of semiclassical Bloch electrons, is given by the expression $D(\bm k) = 1+(e/\hbar)\bm B \cdot \bm\Omega(\bm k)$~\cite{NiuPRL2005}. Here $\bm\Omega(\bm k)= i\langle \nabla_{\bm k}  u (\bm k)| \times | \nabla_{\bm k} u (\bm k) \rangle$ is the Berry curvature, with $\vert u (\bm k)\rangle$ being the Bloch wavefunction of a particular band in the absence of external electromagnetic fields~\cite{BerryReview,Sundaram,NiuPRB1996}. We assume that the spin degeneracy of bands is lifted. If there are multiple bands crossing the chemical potential, the semiclassical motion of electrons in different bands is independent, i.e. there are no interband transitions.

The expression (\ref{Eq:current}) depends on the wavepacket velocity, $\dot{\bm r}$, which can be obtained from the semiclassical equations of motion~\cite{BerryReview,Sundaram,KS2009,Littlejohn}
\begin{equation}
    \dot{\bm r} = \frac{1}{\hbar} \frac{\partial {\tilde\varepsilon}_{\bm k}}{\partial \bm k} - \dot{\bm k} \times \bm \Omega (\bm k),\quad 
    \hbar \dot{\bm k} = -e \bm E - e \dot{\bm r} \times \bm B.
\end{equation}
Here $\tilde{\varepsilon}_{\bm k}= \varepsilon_{\bm k} - \bm m_{\bm k} \cdot \bm B$, $\varepsilon_{\bm k}$ is the Bloch eigenvalue without external fields, and $\bm{m}_{\bm{k}}$ is the total magnetic moment of the wavepacket. Details of calculations are given in the Supplementary Material~\cite{supplementary}. 

Following the Boltzmann transport paradigm, we define a non-equilibrium distribution function $f(\bm k, \bm r, t)$, which satisfies the kinetic equation~\cite{ashcroft1978solid}
\begin{align}
    \frac{\partial f}{\partial t} +\bm {\dot r} \cdot \frac{\partial f}{\partial \bm r} + \bm {\dot k} \cdot \frac{\partial f}{\partial \bm k} 
    = - \frac{f - f_0}{\tau}.
\end{align}
Here $f_0 = 1/[1+e^{\beta(\tilde{\varepsilon} - \mu )}]$ is the equilibrium Fermi-Dirac distribution function, $\mu$ is the chemical potential, and $\beta= 1/k_B T$ is the inverse temperature. We have used the isotropic relaxation time approximation for the collision integral. Assuming a spatially uniform steady state, we drop position and time dependence of the distribution function. 
Treating the applied electric and magnetic fields as perturbations, we solve the Boltzmann equation, assuming $f({\bm k}) = f_0 + \delta f_{\bm k}$~\cite{supplementary}.  

\textbf{Data availability}

The data supporting the findings of this study are available from the 
authors upon reasonable request.


\textbf{Acknowledgments}

This work was supported by the Natural Sciences and Engineering Research Council of Canada through Discovery Grants 2022-05240 (RG) and 2021-03705 (KS).

\textbf{Author contributions}

TF, RG, and KS conceived and designed the research project. TF carried out the calculations. All authors contributed equally to the writing and editing of the manuscript.

\textbf{Competing interests}

The authors declare no competing financial interests.

\bibliography{Ref}

\begin{thebibliography}{39}%
\makeatletter
\providecommand \@ifxundefined [1]{%
 \@ifx{#1\undefined}
}%
\providecommand \@ifnum [1]{%
 \ifnum #1\expandafter \@firstoftwo
 \else \expandafter \@secondoftwo
 \fi
}%
\providecommand \@ifx [1]{%
 \ifx #1\expandafter \@firstoftwo
 \else \expandafter \@secondoftwo
 \fi
}%
\providecommand \natexlab [1]{#1}%
\providecommand \enquote  [1]{``#1''}%
\providecommand \bibnamefont  [1]{#1}%
\providecommand \bibfnamefont [1]{#1}%
\providecommand \citenamefont [1]{#1}%
\providecommand \href@noop [0]{\@secondoftwo}%
\providecommand \href [0]{\begingroup \@sanitize@url \@href}%
\providecommand \@href[1]{\@@startlink{#1}\@@href}%
\providecommand \@@href[1]{\endgroup#1\@@endlink}%
\providecommand \@sanitize@url [0]{\catcode `\\12\catcode `\$12\catcode `\&12\catcode `\#12\catcode `\^12\catcode `\_12\catcode `\%12\relax}%
\providecommand \@@startlink[1]{}%
\providecommand \@@endlink[0]{}%
\providecommand \url  [0]{\begingroup\@sanitize@url \@url }%
\providecommand \@url [1]{\endgroup\@href {#1}{\urlprefix }}%
\providecommand \urlprefix  [0]{URL }%
\providecommand \Eprint [0]{\href }%
\providecommand \doibase [0]{https://doi.org/}%
\providecommand \selectlanguage [0]{\@gobble}%
\providecommand \bibinfo  [0]{\@secondoftwo}%
\providecommand \bibfield  [0]{\@secondoftwo}%
\providecommand \translation [1]{[#1]}%
\providecommand \BibitemOpen [0]{}%
\providecommand \bibitemStop [0]{}%
\providecommand \bibitemNoStop [0]{.\EOS\space}%
\providecommand \EOS [0]{\spacefactor3000\relax}%
\providecommand \BibitemShut  [1]{\csname bibitem#1\endcsname}%
\let\auto@bib@innerbib\@empty
\bibitem [{\citenamefont {Berry}(1984)}]{BerryOriginal}%
  \BibitemOpen
  \bibfield  {author} {\bibinfo {author} {\bibfnamefont {M.~V.}\ \bibnamefont {Berry}},\ }\bibfield  {title} {\bibinfo {title} {Quantal phase factors accompanying adiabatic changes},\ }\href {https://doi.org/10.1098/rspa.1984.0023} {\bibfield  {journal} {\bibinfo  {journal} {Proceedings of the Royal Society of London. A. Mathematical and Physical Sciences}\ }\textbf {\bibinfo {volume} {392}},\ \bibinfo {pages} {45} (\bibinfo {year} {1984})}\BibitemShut {NoStop}%
\bibitem [{\citenamefont {Xiao}\ \emph {et~al.}(2006)\citenamefont {Xiao}, \citenamefont {Yao}, \citenamefont {Fang},\ and\ \citenamefont {Niu}}]{Berry1app}%
  \BibitemOpen
  \bibfield  {author} {\bibinfo {author} {\bibfnamefont {D.}~\bibnamefont {Xiao}}, \bibinfo {author} {\bibfnamefont {Y.}~\bibnamefont {Yao}}, \bibinfo {author} {\bibfnamefont {Z.}~\bibnamefont {Fang}},\ and\ \bibinfo {author} {\bibfnamefont {Q.}~\bibnamefont {Niu}},\ }\bibfield  {title} {\bibinfo {title} {Berry-phase effect in anomalous thermoelectric transport},\ }\href {https://doi.org/10.1103/PhysRevLett.97.026603} {\bibfield  {journal} {\bibinfo  {journal} {Phys. Rev. Lett.}\ }\textbf {\bibinfo {volume} {97}},\ \bibinfo {pages} {026603} (\bibinfo {year} {2006})}\BibitemShut {NoStop}%
\bibitem [{\citenamefont {Moore}\ and\ \citenamefont {Orenstein}(2010)}]{Berry2app}%
  \BibitemOpen
  \bibfield  {author} {\bibinfo {author} {\bibfnamefont {J.~E.}\ \bibnamefont {Moore}}\ and\ \bibinfo {author} {\bibfnamefont {J.}~\bibnamefont {Orenstein}},\ }\bibfield  {title} {\bibinfo {title} {Confinement-induced {B}erry phase and helicity-dependent photocurrents},\ }\href {https://doi.org/10.1103/PhysRevLett.105.026805} {\bibfield  {journal} {\bibinfo  {journal} {Phys. Rev. Lett.}\ }\textbf {\bibinfo {volume} {105}},\ \bibinfo {pages} {026805} (\bibinfo {year} {2010})}\BibitemShut {NoStop}%
\bibitem [{\citenamefont {Qi}\ and\ \citenamefont {Zhang}(2011)}]{Berry3app}%
  \BibitemOpen
  \bibfield  {author} {\bibinfo {author} {\bibfnamefont {X.-L.}\ \bibnamefont {Qi}}\ and\ \bibinfo {author} {\bibfnamefont {S.-C.}\ \bibnamefont {Zhang}},\ }\bibfield  {title} {\bibinfo {title} {Topological insulators and superconductors},\ }\href {https://doi.org/10.1103/RevModPhys.83.1057} {\bibfield  {journal} {\bibinfo  {journal} {Rev. Mod. Phys.}\ }\textbf {\bibinfo {volume} {83}},\ \bibinfo {pages} {1057} (\bibinfo {year} {2011})}\BibitemShut {NoStop}%
\bibitem [{\citenamefont {Shapere}\ and\ \citenamefont {Wilczek}(1989)}]{Berry4app}%
  \BibitemOpen
  \bibfield  {author} {\bibinfo {author} {\bibfnamefont {A.}~\bibnamefont {Shapere}}\ and\ \bibinfo {author} {\bibfnamefont {F.}~\bibnamefont {Wilczek}},\ }\href@noop {} {\emph {\bibinfo {title} {Geometric {P}hases in {P}hysics}}},\ Vol.~\bibinfo {volume} {5}\ (\bibinfo  {publisher} {World scientific},\ \bibinfo {year} {1989})\BibitemShut {NoStop}%
\bibitem [{\citenamefont {Bohm}\ \emph {et~al.}(2013)\citenamefont {Bohm}, \citenamefont {Mostafazadeh}, \citenamefont {Koizumi}, \citenamefont {Niu},\ and\ \citenamefont {Zwanziger}}]{Berry5app}%
  \BibitemOpen
  \bibfield  {author} {\bibinfo {author} {\bibfnamefont {A.}~\bibnamefont {Bohm}}, \bibinfo {author} {\bibfnamefont {A.}~\bibnamefont {Mostafazadeh}}, \bibinfo {author} {\bibfnamefont {H.}~\bibnamefont {Koizumi}}, \bibinfo {author} {\bibfnamefont {Q.}~\bibnamefont {Niu}},\ and\ \bibinfo {author} {\bibfnamefont {J.}~\bibnamefont {Zwanziger}},\ }\href@noop {} {\emph {\bibinfo {title} {The Geometric phase in quantum systems: foundations, mathematical concepts, and applications in molecular and condensed matter physics}}}\ (\bibinfo  {publisher} {Springer Science \& Business Media},\ \bibinfo {year} {2013})\BibitemShut {NoStop}%
\bibitem [{\citenamefont {Haldane}(2004)}]{Berry6app}%
  \BibitemOpen
  \bibfield  {author} {\bibinfo {author} {\bibfnamefont {F.~D.~M.}\ \bibnamefont {Haldane}},\ }\bibfield  {title} {\bibinfo {title} {Berry curvature on the {F}ermi surface: Anomalous {H}all effect as a topological {F}ermi-liquid property},\ }\href {https://doi.org/10.1103/PhysRevLett.93.206602} {\bibfield  {journal} {\bibinfo  {journal} {Phys. Rev. Lett.}\ }\textbf {\bibinfo {volume} {93}},\ \bibinfo {pages} {206602} (\bibinfo {year} {2004})}\BibitemShut {NoStop}%
\bibitem [{\citenamefont {Xiao}\ \emph {et~al.}(2010)\citenamefont {Xiao}, \citenamefont {Chang},\ and\ \citenamefont {Niu}}]{BerryReview}%
  \BibitemOpen
  \bibfield  {author} {\bibinfo {author} {\bibfnamefont {D.}~\bibnamefont {Xiao}}, \bibinfo {author} {\bibfnamefont {M.-C.}\ \bibnamefont {Chang}},\ and\ \bibinfo {author} {\bibfnamefont {Q.}~\bibnamefont {Niu}},\ }\bibfield  {title} {\bibinfo {title} {Berry phase effects on electronic properties},\ }\href {https://doi.org/10.1103/RevModPhys.82.1959} {\bibfield  {journal} {\bibinfo  {journal} {Rev. Mod. Phys.}\ }\textbf {\bibinfo {volume} {82}},\ \bibinfo {pages} {1959} (\bibinfo {year} {2010})}\BibitemShut {NoStop}%
\bibitem [{\citenamefont {Sundaram}\ and\ \citenamefont {Niu}(1999)}]{Sundaram}%
  \BibitemOpen
  \bibfield  {author} {\bibinfo {author} {\bibfnamefont {G.}~\bibnamefont {Sundaram}}\ and\ \bibinfo {author} {\bibfnamefont {Q.}~\bibnamefont {Niu}},\ }\bibfield  {title} {\bibinfo {title} {Wave-packet dynamics in slowly perturbed crystals: Gradient corrections and {Berry}-phase effects},\ }\href {https://doi.org/10.1103/PhysRevB.59.14915} {\bibfield  {journal} {\bibinfo  {journal} {Phys. Rev. B}\ }\textbf {\bibinfo {volume} {59}},\ \bibinfo {pages} {14915} (\bibinfo {year} {1999})}\BibitemShut {NoStop}%
\bibitem [{\citenamefont {Bliokh}(2006{\natexlab{a}})}]{DensityBliokh}%
  \BibitemOpen
  \bibfield  {author} {\bibinfo {author} {\bibfnamefont {K.}~\bibnamefont {Bliokh}},\ }\bibfield  {title} {\bibinfo {title} {On the {H}amiltonian nature of semiclassical equations of motion in the presence of an electromagnetic field and {B}erry curvature},\ }\href {https://doi.org/https://doi.org/10.1016/j.physleta.2005.10.087} {\bibfield  {journal} {\bibinfo  {journal} {Physics Letters A}\ }\textbf {\bibinfo {volume} {351}},\ \bibinfo {pages} {123} (\bibinfo {year} {2006}{\natexlab{a}})}\BibitemShut {NoStop}%
\bibitem [{\citenamefont {Bliokh}(2006{\natexlab{b}})}]{momentBliokh2006}%
  \BibitemOpen
  \bibfield  {author} {\bibinfo {author} {\bibfnamefont {K.~Y.}\ \bibnamefont {Bliokh}},\ }\bibfield  {title} {\bibinfo {title} {Geometrical optics of beams with vortices: Berry phase and orbital angular momentum {H}all effect},\ }\href {https://doi.org/10.1103/PhysRevLett.97.043901} {\bibfield  {journal} {\bibinfo  {journal} {Phys. Rev. Lett.}\ }\textbf {\bibinfo {volume} {97}},\ \bibinfo {pages} {043901} (\bibinfo {year} {2006}{\natexlab{b}})}\BibitemShut {NoStop}%
\bibitem [{\citenamefont {Duval}\ \emph {et~al.}(2006)\citenamefont {Duval}, \citenamefont {Horv\'{a}th}, \citenamefont {Horv\'{a}thy}, \citenamefont {Martina},\ and\ \citenamefont {Stichel}}]{DensityDuval2006}%
  \BibitemOpen
  \bibfield  {author} {\bibinfo {author} {\bibfnamefont {C.}~\bibnamefont {Duval}}, \bibinfo {author} {\bibfnamefont {Z.}~\bibnamefont {Horv\'{a}th}}, \bibinfo {author} {\bibfnamefont {P.~A.}\ \bibnamefont {Horv\'{a}thy}}, \bibinfo {author} {\bibfnamefont {L.}~\bibnamefont {Martina}},\ and\ \bibinfo {author} {\bibfnamefont {P.~C.}\ \bibnamefont {Stichel}},\ }\bibfield  {title} {\bibinfo {title} {{B}erry phase correction to electron densoty in solids and "exotic" dynamics},\ }\href {https://doi.org/10.1142/S0217984906010573} {\bibfield  {journal} {\bibinfo  {journal} {Mod. Phys. Lett. B}\ }\textbf {\bibinfo {volume} {20}},\ \bibinfo {pages} {373} (\bibinfo {year} {2006})}\BibitemShut {NoStop}%
\bibitem [{\citenamefont {Xiao}\ \emph {et~al.}(2007)\citenamefont {Xiao}, \citenamefont {Yao},\ and\ \citenamefont {Niu}}]{NiuPRL2007}%
  \BibitemOpen
  \bibfield  {author} {\bibinfo {author} {\bibfnamefont {D.}~\bibnamefont {Xiao}}, \bibinfo {author} {\bibfnamefont {W.}~\bibnamefont {Yao}},\ and\ \bibinfo {author} {\bibfnamefont {Q.}~\bibnamefont {Niu}},\ }\bibfield  {title} {\bibinfo {title} {Valley-contrasting physics in graphene: Magnetic moment and topological transport},\ }\href {https://doi.org/10.1103/PhysRevLett.99.236809} {\bibfield  {journal} {\bibinfo  {journal} {Phys. Rev. Lett.}\ }\textbf {\bibinfo {volume} {99}},\ \bibinfo {pages} {236809} (\bibinfo {year} {2007})}\BibitemShut {NoStop}%
\bibitem [{\citenamefont {Chang}\ \emph {et~al.}(2023)\citenamefont {Chang}, \citenamefont {Liu},\ and\ \citenamefont {MacDonald}}]{QHEreview2023}%
  \BibitemOpen
  \bibfield  {author} {\bibinfo {author} {\bibfnamefont {C.-Z.}\ \bibnamefont {Chang}}, \bibinfo {author} {\bibfnamefont {C.-X.}\ \bibnamefont {Liu}},\ and\ \bibinfo {author} {\bibfnamefont {A.~H.}\ \bibnamefont {MacDonald}},\ }\bibfield  {title} {\bibinfo {title} {Colloquium: {Q}uantum anomalous {H}all effect},\ }\href {https://doi.org/10.1103/RevModPhys.95.011002} {\bibfield  {journal} {\bibinfo  {journal} {Rev. Mod. Phys.}\ }\textbf {\bibinfo {volume} {95}},\ \bibinfo {pages} {011002} (\bibinfo {year} {2023})}\BibitemShut {NoStop}%
\bibitem [{\citenamefont {Sodemann}\ and\ \citenamefont {Fu}(2015)}]{LiangFu2015}%
  \BibitemOpen
  \bibfield  {author} {\bibinfo {author} {\bibfnamefont {I.}~\bibnamefont {Sodemann}}\ and\ \bibinfo {author} {\bibfnamefont {L.}~\bibnamefont {Fu}},\ }\bibfield  {title} {\bibinfo {title} {Quantum nonlinear {H}all effect induced by {B}erry curvature dipole in time-reversal invariant materials},\ }\href {https://doi.org/10.1103/PhysRevLett.115.216806} {\bibfield  {journal} {\bibinfo  {journal} {Phys. Rev. Lett.}\ }\textbf {\bibinfo {volume} {115}},\ \bibinfo {pages} {216806} (\bibinfo {year} {2015})}\BibitemShut {NoStop}%
\bibitem [{\citenamefont {Zhang}\ \emph {et~al.}(2023)\citenamefont {Zhang}, \citenamefont {Gao}, \citenamefont {Xie}, \citenamefont {Po},\ and\ \citenamefont {Law}}]{LawPRB2023}%
  \BibitemOpen
  \bibfield  {author} {\bibinfo {author} {\bibfnamefont {C.-P.}\ \bibnamefont {Zhang}}, \bibinfo {author} {\bibfnamefont {X.-J.}\ \bibnamefont {Gao}}, \bibinfo {author} {\bibfnamefont {Y.-M.}\ \bibnamefont {Xie}}, \bibinfo {author} {\bibfnamefont {H.~C.}\ \bibnamefont {Po}},\ and\ \bibinfo {author} {\bibfnamefont {K.~T.}\ \bibnamefont {Law}},\ }\bibfield  {title} {\bibinfo {title} {Higher-order nonlinear anomalous {H}all effects induced by {B}erry curvature multipoles},\ }\href {https://doi.org/10.1103/PhysRevB.107.115142} {\bibfield  {journal} {\bibinfo  {journal} {Phys. Rev. B}\ }\textbf {\bibinfo {volume} {107}},\ \bibinfo {pages} {115142} (\bibinfo {year} {2023})}\BibitemShut {NoStop}%
\bibitem [{\citenamefont {Farajollahpour}\ \emph {et~al.}(2025)\citenamefont {Farajollahpour}, \citenamefont {Ganesh},\ and\ \citenamefont {Samokhin}}]{Tohid2024}%
  \BibitemOpen
  \bibfield  {author} {\bibinfo {author} {\bibfnamefont {T.}~\bibnamefont {Farajollahpour}}, \bibinfo {author} {\bibfnamefont {R.}~\bibnamefont {Ganesh}},\ and\ \bibinfo {author} {\bibfnamefont {K.~V.}\ \bibnamefont {Samokhin}},\ }\bibfield  {title} {\bibinfo {title} {Light-induced charge and spin {H}all currents in materials with $c_4k$ symmetry},\ }\href {https://doi.org/10.1038/s41535-025-00746-7} {\bibfield  {journal} {\bibinfo  {journal} {npj Quantum Materials}\ }\textbf {\bibinfo {volume} {10}},\ \bibinfo {pages} {29} (\bibinfo {year} {2025})}\BibitemShut {NoStop}%
\bibitem [{\citenamefont {\ifmmode~\check{S}\else \v{S}\fi{}mejkal}\ \emph {et~al.}(2022{\natexlab{a}})\citenamefont {\ifmmode~\check{S}\else \v{S}\fi{}mejkal}, \citenamefont {Sinova},\ and\ \citenamefont {Jungwirth}}]{SinovaPRX2022}%
  \BibitemOpen
  \bibfield  {author} {\bibinfo {author} {\bibfnamefont {L.}~\bibnamefont {\ifmmode~\check{S}\else \v{S}\fi{}mejkal}}, \bibinfo {author} {\bibfnamefont {J.}~\bibnamefont {Sinova}},\ and\ \bibinfo {author} {\bibfnamefont {T.}~\bibnamefont {Jungwirth}},\ }\bibfield  {title} {\bibinfo {title} {Emerging research landscape of altermagnetism},\ }\href {https://doi.org/10.1103/PhysRevX.12.040501} {\bibfield  {journal} {\bibinfo  {journal} {Phys. Rev. X}\ }\textbf {\bibinfo {volume} {12}},\ \bibinfo {pages} {040501} (\bibinfo {year} {2022}{\natexlab{a}})}\BibitemShut {NoStop}%
\bibitem [{\citenamefont {\ifmmode~\check{S}\else \v{S}\fi{}mejkal}\ \emph {et~al.}(2022{\natexlab{b}})\citenamefont {\ifmmode~\check{S}\else \v{S}\fi{}mejkal}, \citenamefont {Sinova},\ and\ \citenamefont {Jungwirth}}]{Sinova2PRX2022}%
  \BibitemOpen
  \bibfield  {author} {\bibinfo {author} {\bibfnamefont {L.}~\bibnamefont {\ifmmode~\check{S}\else \v{S}\fi{}mejkal}}, \bibinfo {author} {\bibfnamefont {J.}~\bibnamefont {Sinova}},\ and\ \bibinfo {author} {\bibfnamefont {T.}~\bibnamefont {Jungwirth}},\ }\bibfield  {title} {\bibinfo {title} {Beyond conventional ferromagnetism and antiferromagnetism: A phase with nonrelativistic spin and crystal rotation symmetry},\ }\href {https://doi.org/10.1103/PhysRevX.12.031042} {\bibfield  {journal} {\bibinfo  {journal} {Phys. Rev. X}\ }\textbf {\bibinfo {volume} {12}},\ \bibinfo {pages} {031042} (\bibinfo {year} {2022}{\natexlab{b}})}\BibitemShut {NoStop}%
\bibitem [{\citenamefont {Fedchenko}\ \emph {et~al.}(2024)\citenamefont {Fedchenko}, \citenamefont {Min{\'a}r}, \citenamefont {Akashdeep}, \citenamefont {D'Souza}, \citenamefont {Vasilyev}, \citenamefont {Tkach}, \citenamefont {Odenbreit}, \citenamefont {Nguyen}, \citenamefont {Kutnyakhov}, \citenamefont {Wind}, \citenamefont {Wenthaus}, \citenamefont {Scholz}, \citenamefont {Rossnagel}, \citenamefont {Hoesch}, \citenamefont {Aeschlimann}, \citenamefont {Stadtm{\"u}ller}, \citenamefont {Kl{\"a}ui}, \citenamefont {Sch{\"o}nhense}, \citenamefont {Jungwirth}, \citenamefont {Hellenes}, \citenamefont {Jakob}, \citenamefont {{\v S}mejkal}, \citenamefont {Sinova},\ and\ \citenamefont {Elmers}}]{Fedchenko_undated-qm}%
  \BibitemOpen
  \bibfield  {author} {\bibinfo {author} {\bibfnamefont {O.}~\bibnamefont {Fedchenko}}, \bibinfo {author} {\bibfnamefont {J.}~\bibnamefont {Min{\'a}r}}, \bibinfo {author} {\bibfnamefont {A.}~\bibnamefont {Akashdeep}}, \bibinfo {author} {\bibfnamefont {S.~W.}\ \bibnamefont {D'Souza}}, \bibinfo {author} {\bibfnamefont {D.}~\bibnamefont {Vasilyev}}, \bibinfo {author} {\bibfnamefont {O.}~\bibnamefont {Tkach}}, \bibinfo {author} {\bibfnamefont {L.}~\bibnamefont {Odenbreit}}, \bibinfo {author} {\bibfnamefont {Q.}~\bibnamefont {Nguyen}}, \bibinfo {author} {\bibfnamefont {D.}~\bibnamefont {Kutnyakhov}}, \bibinfo {author} {\bibfnamefont {N.}~\bibnamefont {Wind}}, \bibinfo {author} {\bibfnamefont {L.}~\bibnamefont {Wenthaus}}, \bibinfo {author} {\bibfnamefont {M.}~\bibnamefont {Scholz}}, \bibinfo {author} {\bibfnamefont {K.}~\bibnamefont {Rossnagel}}, \bibinfo {author} {\bibfnamefont {M.}~\bibnamefont {Hoesch}}, \bibinfo {author} {\bibfnamefont {M.}~\bibnamefont {Aeschlimann}}, \bibinfo {author} {\bibfnamefont
  {B.}~\bibnamefont {Stadtm{\"u}ller}}, \bibinfo {author} {\bibfnamefont {M.}~\bibnamefont {Kl{\"a}ui}}, \bibinfo {author} {\bibfnamefont {G.}~\bibnamefont {Sch{\"o}nhense}}, \bibinfo {author} {\bibfnamefont {T.}~\bibnamefont {Jungwirth}}, \bibinfo {author} {\bibfnamefont {A.~B.}\ \bibnamefont {Hellenes}}, \bibinfo {author} {\bibfnamefont {G.}~\bibnamefont {Jakob}}, \bibinfo {author} {\bibfnamefont {L.}~\bibnamefont {{\v S}mejkal}}, \bibinfo {author} {\bibfnamefont {J.}~\bibnamefont {Sinova}},\ and\ \bibinfo {author} {\bibfnamefont {H.-J.}\ \bibnamefont {Elmers}},\ }\bibfield  {title} {\bibinfo {title} {Observation of time-reversal symmetry breaking in the band structure of altermagnetic {R}u{O}$_2$},\ }\href {https://doi.org/DOI: 10.1126/sciadv.adj4883} {\bibfield  {journal} {\bibinfo  {journal} {Science Advances}\ }\textbf {\bibinfo {volume} {10}},\ \bibinfo {pages} {4883} (\bibinfo {year} {2024})}\BibitemShut {NoStop}%
\bibitem [{\citenamefont {Schindler}\ \emph {et~al.}(2018)\citenamefont {Schindler}, \citenamefont {Cook}, \citenamefont {Vergniory}, \citenamefont {Wang}, \citenamefont {Parkin}, \citenamefont {Bernevig},\ and\ \citenamefont {Neupert}}]{Schindler2018}%
  \BibitemOpen
  \bibfield  {author} {\bibinfo {author} {\bibfnamefont {F.}~\bibnamefont {Schindler}}, \bibinfo {author} {\bibfnamefont {A.~M.}\ \bibnamefont {Cook}}, \bibinfo {author} {\bibfnamefont {M.~G.}\ \bibnamefont {Vergniory}}, \bibinfo {author} {\bibfnamefont {Z.}~\bibnamefont {Wang}}, \bibinfo {author} {\bibfnamefont {S.~S.~P.}\ \bibnamefont {Parkin}}, \bibinfo {author} {\bibfnamefont {B.~A.}\ \bibnamefont {Bernevig}},\ and\ \bibinfo {author} {\bibfnamefont {T.}~\bibnamefont {Neupert}},\ }\bibfield  {title} {\bibinfo {title} {Higher-order topological insulators},\ }\href {https://doi.org/10.1126/sciadv.aat0346} {\bibfield  {journal} {\bibinfo  {journal} {Science Advances}\ }\textbf {\bibinfo {volume} {4}},\ \bibinfo {pages} {0346} (\bibinfo {year} {2018})}\BibitemShut {NoStop}%
\bibitem [{\citenamefont {Wang}\ \emph {et~al.}(2007)\citenamefont {Wang}, \citenamefont {Vanderbilt}, \citenamefont {Yates},\ and\ \citenamefont {Souza}}]{Wang2007}%
  \BibitemOpen
  \bibfield  {author} {\bibinfo {author} {\bibfnamefont {X.}~\bibnamefont {Wang}}, \bibinfo {author} {\bibfnamefont {D.}~\bibnamefont {Vanderbilt}}, \bibinfo {author} {\bibfnamefont {J.~R.}\ \bibnamefont {Yates}},\ and\ \bibinfo {author} {\bibfnamefont {I.}~\bibnamefont {Souza}},\ }\bibfield  {title} {\bibinfo {title} {Fermi-surface calculation of the anomalous {H}all conductivity},\ }\href {https://doi.org/10.1103/PhysRevB.76.195109} {\bibfield  {journal} {\bibinfo  {journal} {Phys. Rev. B}\ }\textbf {\bibinfo {volume} {76}},\ \bibinfo {pages} {195109} (\bibinfo {year} {2007})}\BibitemShut {NoStop}%
\bibitem [{\citenamefont {Hurd}(2012)}]{hurd2012hall}%
  \BibitemOpen
  \bibfield  {author} {\bibinfo {author} {\bibfnamefont {C.}~\bibnamefont {Hurd}},\ }\href@noop {} {\emph {\bibinfo {title} {The {H}all {E}ffect in {M}etals and {A}lloys}}}\ (\bibinfo  {publisher} {Springer Science \& Business Media},\ \bibinfo {year} {2012})\BibitemShut {NoStop}%
\bibitem [{\citenamefont {Ziman}(1979)}]{Ziman1979}%
  \BibitemOpen
  \bibfield  {author} {\bibinfo {author} {\bibfnamefont {J.~M.}\ \bibnamefont {Ziman}},\ }\href@noop {} {\emph {\bibinfo {title} {Principles of the Theory of Solids}}}\ (\bibinfo  {publisher} {Cambridge university press},\ \bibinfo {year} {1979})\BibitemShut {NoStop}%
\bibitem [{\citenamefont {Nagaosa}\ \emph {et~al.}(2010{\natexlab{a}})\citenamefont {Nagaosa}, \citenamefont {Sinova}, \citenamefont {Onoda}, \citenamefont {MacDonald},\ and\ \citenamefont {Ong}}]{Nagaosa2010}%
  \BibitemOpen
  \bibfield  {author} {\bibinfo {author} {\bibfnamefont {N.}~\bibnamefont {Nagaosa}}, \bibinfo {author} {\bibfnamefont {J.}~\bibnamefont {Sinova}}, \bibinfo {author} {\bibfnamefont {S.}~\bibnamefont {Onoda}}, \bibinfo {author} {\bibfnamefont {A.~H.}\ \bibnamefont {MacDonald}},\ and\ \bibinfo {author} {\bibfnamefont {N.~P.}\ \bibnamefont {Ong}},\ }\bibfield  {title} {\bibinfo {title} {Anomalous {H}all effect},\ }\href {https://doi.org/10.1103/RevModPhys.82.1539} {\bibfield  {journal} {\bibinfo  {journal} {Rev. Mod. Phys.}\ }\textbf {\bibinfo {volume} {82}},\ \bibinfo {pages} {1539} (\bibinfo {year} {2010}{\natexlab{a}})}\BibitemShut {NoStop}%
\bibitem [{sup()}]{supplementary}%
  \BibitemOpen
  \href@noop {} {\bibinfo  {journal} {Supplementary~material}\ }\BibitemShut {NoStop}%
\bibitem [{\citenamefont {Landau}\ and\ \citenamefont {Lifshitz}(2001)}]{LL-5}%
  \BibitemOpen
\bibfield  {journal} {  }\bibfield  {author} {\bibinfo {author} {\bibfnamefont {L.~D.}\ \bibnamefont {Landau}}\ and\ \bibinfo {author} {\bibfnamefont {E.~M.}\ \bibnamefont {Lifshitz}},\ }\href@noop {} {\emph {\bibinfo {title} {Statistical {P}hysics, {P}art 1}}}\ (\bibinfo  {publisher} {Butterworth-Heinemann},\ \bibinfo {year} {2001})\BibitemShut {NoStop}%
\bibitem [{\citenamefont {Occhialini}\ \emph {et~al.}(2021)\citenamefont {Occhialini}, \citenamefont {Bisogni}, \citenamefont {You}, \citenamefont {Barbour}, \citenamefont {Jarrige}, \citenamefont {Mitchell}, \citenamefont {Comin},\ and\ \citenamefont {Pelliciari}}]{Occhialini2021}%
  \BibitemOpen
  \bibfield  {author} {\bibinfo {author} {\bibfnamefont {C.~A.}\ \bibnamefont {Occhialini}}, \bibinfo {author} {\bibfnamefont {V.}~\bibnamefont {Bisogni}}, \bibinfo {author} {\bibfnamefont {H.}~\bibnamefont {You}}, \bibinfo {author} {\bibfnamefont {A.}~\bibnamefont {Barbour}}, \bibinfo {author} {\bibfnamefont {I.}~\bibnamefont {Jarrige}}, \bibinfo {author} {\bibfnamefont {J.~F.}\ \bibnamefont {Mitchell}}, \bibinfo {author} {\bibfnamefont {R.}~\bibnamefont {Comin}},\ and\ \bibinfo {author} {\bibfnamefont {J.}~\bibnamefont {Pelliciari}},\ }\bibfield  {title} {\bibinfo {title} {Local electronic structure of rutile {R}u{O}$_2$},\ }\href {https://doi.org/10.1103/PhysRevResearch.3.033214} {\bibfield  {journal} {\bibinfo  {journal} {Phys. Rev. Res.}\ }\textbf {\bibinfo {volume} {3}},\ \bibinfo {pages} {033214} (\bibinfo {year} {2021})}\BibitemShut {NoStop}%
\bibitem [{\citenamefont {Jiang}\ \emph {et~al.}(2025)\citenamefont {Jiang}, \citenamefont {Hu}, \citenamefont {Bai}, \citenamefont {Song}, \citenamefont {Mu}, \citenamefont {Qu}, \citenamefont {Li}, \citenamefont {Zhu}, \citenamefont {Pi}, \citenamefont {Wei}, \citenamefont {Sun}, \citenamefont {Huang}, \citenamefont {Zheng}, \citenamefont {Peng}, \citenamefont {He}, \citenamefont {Li}, \citenamefont {Luo}, \citenamefont {Li}, \citenamefont {Chen}, \citenamefont {Li}, \citenamefont {Weng},\ and\ \citenamefont {Qian}}]{Jiang2025}%
  \BibitemOpen
  \bibfield  {author} {\bibinfo {author} {\bibfnamefont {B.}~\bibnamefont {Jiang}}, \bibinfo {author} {\bibfnamefont {M.}~\bibnamefont {Hu}}, \bibinfo {author} {\bibfnamefont {J.}~\bibnamefont {Bai}}, \bibinfo {author} {\bibfnamefont {Z.}~\bibnamefont {Song}}, \bibinfo {author} {\bibfnamefont {C.}~\bibnamefont {Mu}}, \bibinfo {author} {\bibfnamefont {G.}~\bibnamefont {Qu}}, \bibinfo {author} {\bibfnamefont {W.}~\bibnamefont {Li}}, \bibinfo {author} {\bibfnamefont {W.}~\bibnamefont {Zhu}}, \bibinfo {author} {\bibfnamefont {H.}~\bibnamefont {Pi}}, \bibinfo {author} {\bibfnamefont {Z.}~\bibnamefont {Wei}}, \bibinfo {author} {\bibfnamefont {Y.-J.}\ \bibnamefont {Sun}}, \bibinfo {author} {\bibfnamefont {Y.}~\bibnamefont {Huang}}, \bibinfo {author} {\bibfnamefont {X.}~\bibnamefont {Zheng}}, \bibinfo {author} {\bibfnamefont {Y.}~\bibnamefont {Peng}}, \bibinfo {author} {\bibfnamefont {L.}~\bibnamefont {He}}, \bibinfo {author} {\bibfnamefont {S.}~\bibnamefont {Li}}, \bibinfo {author} {\bibfnamefont {J.}~\bibnamefont
  {Luo}}, \bibinfo {author} {\bibfnamefont {Z.}~\bibnamefont {Li}}, \bibinfo {author} {\bibfnamefont {G.}~\bibnamefont {Chen}}, \bibinfo {author} {\bibfnamefont {H.}~\bibnamefont {Li}}, \bibinfo {author} {\bibfnamefont {H.}~\bibnamefont {Weng}},\ and\ \bibinfo {author} {\bibfnamefont {T.}~\bibnamefont {Qian}},\ }\bibfield  {title} {\bibinfo {title} {A metallic room-temperature d-wave altermagnet},\ }\href {https://doi.org/https://doi.org/10.1038/s41567-025-02822-y} {\bibfield  {journal} {\bibinfo  {journal} {Nature}\ ,\ \bibinfo {pages} {645}} (\bibinfo {year} {2025})}\BibitemShut {NoStop}%
\bibitem [{\citenamefont {Xiao}\ \emph {et~al.}(2005)\citenamefont {Xiao}, \citenamefont {Shi},\ and\ \citenamefont {Niu}}]{NiuPRL2005}%
  \BibitemOpen
  \bibfield  {author} {\bibinfo {author} {\bibfnamefont {D.}~\bibnamefont {Xiao}}, \bibinfo {author} {\bibfnamefont {J.}~\bibnamefont {Shi}},\ and\ \bibinfo {author} {\bibfnamefont {Q.}~\bibnamefont {Niu}},\ }\bibfield  {title} {\bibinfo {title} {Berry phase correction to electron density of states in solids},\ }\href {https://doi.org/10.1103/PhysRevLett.95.137204} {\bibfield  {journal} {\bibinfo  {journal} {Phys. Rev. Lett.}\ }\textbf {\bibinfo {volume} {95}},\ \bibinfo {pages} {137204} (\bibinfo {year} {2005})}\BibitemShut {NoStop}%
\bibitem [{\citenamefont {Chang}\ and\ \citenamefont {Niu}(1996)}]{NiuPRB1996}%
  \BibitemOpen
  \bibfield  {author} {\bibinfo {author} {\bibfnamefont {M.-C.}\ \bibnamefont {Chang}}\ and\ \bibinfo {author} {\bibfnamefont {Q.}~\bibnamefont {Niu}},\ }\bibfield  {title} {\bibinfo {title} {Berry phase, hyperorbits, and the {H}ofstadter spectrum: Semiclassical dynamics in magnetic {B}loch bands},\ }\href {https://doi.org/10.1103/PhysRevB.53.7010} {\bibfield  {journal} {\bibinfo  {journal} {Phys. Rev. B}\ }\textbf {\bibinfo {volume} {53}},\ \bibinfo {pages} {7010} (\bibinfo {year} {1996})}\BibitemShut {NoStop}%
\bibitem [{\citenamefont {Samokhin}(2009)}]{KS2009}%
  \BibitemOpen
  \bibfield  {author} {\bibinfo {author} {\bibfnamefont {K.}~\bibnamefont {Samokhin}},\ }\bibfield  {title} {\bibinfo {title} {Spin–orbit coupling and semiclassical electron dynamics in noncentrosymmetric metals},\ }\href {https://doi.org/https://doi.org/10.1016/j.aop.2009.08.008} {\bibfield  {journal} {\bibinfo  {journal} {Annals of Physics}\ }\textbf {\bibinfo {volume} {324}},\ \bibinfo {pages} {2385} (\bibinfo {year} {2009})}\BibitemShut {NoStop}%
\bibitem [{\citenamefont {Littlejohn}\ and\ \citenamefont {Flynn}(1991)}]{Littlejohn}%
  \BibitemOpen
  \bibfield  {author} {\bibinfo {author} {\bibfnamefont {R.~G.}\ \bibnamefont {Littlejohn}}\ and\ \bibinfo {author} {\bibfnamefont {W.~G.}\ \bibnamefont {Flynn}},\ }\bibfield  {title} {\bibinfo {title} {Geometric phases in the asymptotic theory of coupled wave equations},\ }\href {https://doi.org/10.1103/PhysRevA.44.5239} {\bibfield  {journal} {\bibinfo  {journal} {Phys. Rev. A}\ }\textbf {\bibinfo {volume} {44}},\ \bibinfo {pages} {5239} (\bibinfo {year} {1991})}\BibitemShut {NoStop}%
\bibitem [{\citenamefont {Ashcroft}\ and\ \citenamefont {Mermin}(1976)}]{ashcroft1978solid}%
  \BibitemOpen
  \bibfield  {author} {\bibinfo {author} {\bibfnamefont {N.~W.}\ \bibnamefont {Ashcroft}}\ and\ \bibinfo {author} {\bibfnamefont {N.~D.}\ \bibnamefont {Mermin}},\ }\href@noop {} {\emph {\bibinfo {title} {Solid {S}tate {P}hysics}}}\ (\bibinfo  {publisher} {Holt, Rinehart and Winston},\ \bibinfo {year} {1976})\BibitemShut {NoStop}%
\bibitem [{\citenamefont {Nagaosa}\ \emph {et~al.}(2010{\natexlab{b}})\citenamefont {Nagaosa}, \citenamefont {Sinova}, \citenamefont {Onoda}, \citenamefont {MacDonald},\ and\ \citenamefont {Ong}}]{AnomalRev2010}%
  \BibitemOpen
  \bibfield  {author} {\bibinfo {author} {\bibfnamefont {N.}~\bibnamefont {Nagaosa}}, \bibinfo {author} {\bibfnamefont {J.}~\bibnamefont {Sinova}}, \bibinfo {author} {\bibfnamefont {S.}~\bibnamefont {Onoda}}, \bibinfo {author} {\bibfnamefont {A.~H.}\ \bibnamefont {MacDonald}},\ and\ \bibinfo {author} {\bibfnamefont {N.~P.}\ \bibnamefont {Ong}},\ }\bibfield  {title} {\bibinfo {title} {Anomalous {H}all effect},\ }\href {https://doi.org/10.1103/RevModPhys.82.1539} {\bibfield  {journal} {\bibinfo  {journal} {Rev. Mod. Phys.}\ }\textbf {\bibinfo {volume} {82}},\ \bibinfo {pages} {1539} (\bibinfo {year} {2010}{\natexlab{b}})}\BibitemShut {NoStop}%
\bibitem [{\citenamefont {Shen}(2018)}]{shun2018topological}%
  \BibitemOpen
  \bibfield  {author} {\bibinfo {author} {\bibfnamefont {S.-Q.}\ \bibnamefont {Shen}},\ }\href {https://doi.org/https://doi.org/10.1007/978-3-642-32858-9} {\emph {\bibinfo {title} {Topological Insulators: Dirac Equation in Condensed Matter}}}\ (\bibinfo  {publisher} {Springer Berlin},\ \bibinfo {year} {2018})\BibitemShut {NoStop}%
\bibitem [{\citenamefont {Panati}\ \emph {et~al.}(2003)\citenamefont {Panati}, \citenamefont {Spohn},\ and\ \citenamefont {Teufel}}]{Panati2003}%
  \BibitemOpen
  \bibfield  {author} {\bibinfo {author} {\bibfnamefont {G.}~\bibnamefont {Panati}}, \bibinfo {author} {\bibfnamefont {H.}~\bibnamefont {Spohn}},\ and\ \bibinfo {author} {\bibfnamefont {S.}~\bibnamefont {Teufel}},\ }\bibfield  {title} {\bibinfo {title} {Effective dynamics for bloch electrons: Peierls substitution and beyond},\ }\href {https://doi.org/10.1007/s00220-003-0950-1} {\bibfield  {journal} {\bibinfo  {journal} {Communications in Mathematical Physics}\ }\textbf {\bibinfo {volume} {242}},\ \bibinfo {pages} {547} (\bibinfo {year} {2003})}\BibitemShut {NoStop}%
\bibitem [{\citenamefont {Yang}\ \emph {et~al.}(2023)\citenamefont {Yang}, \citenamefont {Li}, \citenamefont {Luo}, \citenamefont {Miao}, \citenamefont {Chen},\ and\ \citenamefont {Xing}}]{YangPRB2023}%
  \BibitemOpen
  \bibfield  {author} {\bibinfo {author} {\bibfnamefont {M.-X.}\ \bibnamefont {Yang}}, \bibinfo {author} {\bibfnamefont {H.-D.}\ \bibnamefont {Li}}, \bibinfo {author} {\bibfnamefont {W.}~\bibnamefont {Luo}}, \bibinfo {author} {\bibfnamefont {B.}~\bibnamefont {Miao}}, \bibinfo {author} {\bibfnamefont {W.}~\bibnamefont {Chen}},\ and\ \bibinfo {author} {\bibfnamefont {D.~Y.}\ \bibnamefont {Xing}},\ }\bibfield  {title} {\bibinfo {title} {Topological linear magnetoresistivity and thermoconductivity induced by noncentrosymmetric {B}erry curvature},\ }\href {https://doi.org/10.1103/PhysRevB.107.165130} {\bibfield  {journal} {\bibinfo  {journal} {Phys. Rev. B}\ }\textbf {\bibinfo {volume} {107}},\ \bibinfo {pages} {165130} (\bibinfo {year} {2023})}\BibitemShut {NoStop}%
\bibitem [{\citenamefont {Abrikosov}(2017)}]{Abr-book}%
  \BibitemOpen
  \bibfield  {author} {\bibinfo {author} {\bibfnamefont {A.~A.}\ \bibnamefont {Abrikosov}},\ }\href@noop {} {\emph {\bibinfo {title} {Fundamentals of the Theory of Metals}}}\ (\bibinfo  {publisher} {Dover Publications},\ \bibinfo {year} {2017})\BibitemShut {NoStop}%
\end{thebibliography}%

\newpage \clearpage
 
\onecolumngrid 
\setcounter{secnumdepth}{3}
\renewcommand{\theequation}{S\arabic{equation}}
\renewcommand{\thefigure}{S\arabic{figure}}
\begin{center}
 \textbf{\large Supplemental Material for ``Berry curvature-induced transport signature for altermagnetic order''}\\[.5cm]
T. Farajollahpour,$^1$ R. Ganesh,$^1$ and K. V. Samokhin$^1$\\[.4cm]
{\itshape ${}^1$Department of Physics, Brock University, St. Catharines, Ontario L2S 3A1, Canada\\}
(Dated: \today)\\[2cm]
\end{center}

\section{Conductivity from semiclassical equations of motion}

We assume non-degenerate bands, with spin degeneracy lifted due to broken time-reversal symmetry. If multiple bands cross the chemical potential, the motion of electrons in each band is taken to be independent, i.e., there are no interband transitions. In the semiclassical picture, we describe electrons as forming a finite-sized wavepacket, with average position $\bm r$ and average wavevector $\bm k$. Dynamics of the wavepacket is described by semiclassical equations of motion modified by the Berry curvature~\cite{BerryReview,AnomalRev2010,Berry5app,shun2018topological,Littlejohn}: 
\begin{align}
\dot{\bm r} = \frac{1}{\hbar} \frac{\partial \tilde{\varepsilon}_{\bm k}}{\partial {\bm k}} - \dot{\bm k} \times \bm \Omega(\bm k),
\label{eq:EOM-1}
\end{align}
and
\begin{align}
 \hbar \dot{\bm k} = -e \bm E - e \dot{\bm r} \times \bm B,
\label{eq:EOM-2}
\end{align}
where $\bm E$ and $\bm B$ are the external electromagnetic fields and $\tilde{\varepsilon}_{\bm k}$ represents the energy of the Bloch state. In the absence of external fields, the latter is the same as the band energy which we denote as $\varepsilon_{\bm k}$. In the presence of an external magnetic field, it takes the form $\tilde{\varepsilon}_{\bm k }= \varepsilon_{\bm k} -{ \bm m}_{\bm k} \cdot {\bm B}$. Here $\bm m_{\bm k}$ is the magnetic moment of the wavepacket, which has two contributions: $\bm{m}^{\mathrm{orb}}_{\bm{k}}$, an orbital moment due to the wavepacket's self-rotation, and $\bm{m}^{\mathrm{s}}_{\bm{k}}$, an intrinsic spin moment~\cite{BerryReview,Sundaram,NiuPRB1996,Panati2003,KS2009}.  

The steady-state electric current density in a spatially uniform system can be expressed in 
terms of the distribution function $f({\bm k})$ as 
\begin{align}
     j_i = -e \int \frac{d^d{\bm k}}{(2\pi)^d}\, D(\bm k)\, {\dot{r}}_i\, f(\bm k), 
     \label{Eq:current}
\end{align}
where $d=2$ or $3$ is the dimensionality of the material and the electron charge is $-e$. The measure $D(\bm k)$, which encodes a modification of the phase space density arising from the noncanonical dynamics of semiclassical Bloch electrons, is given by $D(\bm k) = 1+(e/\hbar)~\bm B \cdot \bm \Omega(\bm k)$~\cite{NiuPRL2005}. Here, ${\bm \Omega}(\bm k)= i\langle {\bm \nabla}_{\bm k}  u (\bm k)| \times | {\bm\nabla}_{\bm k} u (\bm k) \rangle$ is the Berry curvature, with $\vert u (\bm k)\rangle$ being the Bloch wavefunction in a particular band in the absence of external electromagnetic fields
~\cite{BerryReview,Sundaram,NiuPRB1996}. Note that for a two-dimensional sample in the $xy$ plane, $\bm \Omega$ has only one non-zero component (along $z$). The electron distribution function is obtained from the Boltzmann equation
\begin{align}
    \bm {\dot k} \cdot \frac{\partial f(\bm k)}{\partial {\bm k}} =\frac{{\rm d}f}{{\rm d}t}\bigg|_{\rm collision}= - \frac{f(\bm k) - f_0}{\tau},
\end{align}
where we have invoked the relaxation time approximation for the collision integral. We consider a single relaxation time for all bands, with no dependence on momentum. 

We seek to find the correction to the distribution function due to externally applied fields. Writing $f(\bm k) = f_0 + \delta f(\bm k)$, we solve the Boltzmann equation and calculate $\delta f(\bm k)$ in an order-by-order fashion, 
\begin{align}
    \delta f(\bm k) = \sum_{n=1}^\infty (-\tau \dot{\bm k}\cdot {\bm \partial_k})^n f_0(\tilde{\varepsilon}),
\end{align}
where the $n^{\mathrm{th}}$ order term is the $\mathcal{O}(\tau^n)$ correction. Retaining terms up to second order, we have
\begin{align}
    f(\bm k) = f_0(\tilde{\varepsilon}) + \delta f_1 + \delta f_2, 
\end{align}
where $f_0(\tilde{\varepsilon})$ is the equilibrium Fermi distribution function in the presence of a magnetic field, given by
$f_0(\tilde{\varepsilon}) = f_0(\varepsilon - {\bm m\cdot \bm B}) \simeq f_0(\varepsilon) - f'_0(\varepsilon){\bm m\cdot \bm B}$ to linear order in $\bm{B}$. For the corrections, we obtain the following expressions:
\begin{align}
   & \delta f_1 = -\tau \dot{k}_i \frac{\partial f_0}{\partial k_i}, \nonumber\\
   & \delta f_2 = -\tau \dot{k}_i\frac{\partial f_1}{\partial k_i} = \tau^2  \left(\dot{k}_i \frac{\partial \dot{k}_j }{\partial k_i}  \frac{\partial f_0}{\partial k_j} + \dot{k}_i\dot{k}_j \frac{\partial^2 f_0}{\partial k_i \partial k_j}\right),
    \label{Eq:f1f2}
\end{align}
where $\dot{\bm k}$ is given by
\begin{align}
    \dot{\bm k} = D^{-1} \left[ -\frac{e}{\hbar}\bm{E} -\frac{e}{\hbar} ({\tilde{\bm v}}\times \bm B) - \frac{e^2}{\hbar^2} (\bm E \cdot \bm B)\, \bm\Omega\right], 
    \label{Eq:kdot}
\end{align}
with 
$$
    \tilde{\bm v} = {\frac{1}{\hbar}} \frac{\partial \tilde{\varepsilon}_{\bm k}}{\partial {\bm k}}= {\bm v} - {\frac{1}{\hbar}}\frac{\partial}{\partial\bm{k}} (\bm m \cdot \bm{B})
$$
being the group velocity derived from $\tilde{\varepsilon}_{\bm k}$. To obtain Eq.~(\ref{Eq:kdot}), we substitute Eq.~(\ref{eq:EOM-1}) into Eq.~(\ref{eq:EOM-2}), which yields
\begin{align}
\dot{\bm{k}} = -\frac{e}{\hbar} \bm{E} - \frac{e}{\hbar} \tilde{\bm{v}} \times \bm{B} + \frac{e}{\hbar} (\dot{\bm{k}} \times \bm{\Omega}) \times \bm{B}=-\frac{e}{\hbar} \bm{E} - \frac{e}{\hbar} \tilde{\bm{v}} \times \bm{B} - \frac{e}{\hbar} \dot{\bm{k}}(\bm{B}\cdot\bm{\Omega})+\frac{e}{\hbar} \bm{\Omega}(\dot{\bm{k}}\cdot\bm{B}).
\label{kdot1}
\end{align}
Substituting Eq.~(\ref{eq:EOM-2}) in the last term and recognizing that $(\dot{\bm r} \times \bm B)\cdot \bm B=0$, we arrive at Eq.~(\ref{Eq:kdot}). 

To evaluate Eq.~(\ref{Eq:current}), we follow an analogous procedure to find $ \dot{\bm r}$. We substitute $\dot{\bm k}$ into $\dot{\bm r}$ and obtain:
\begin{align}
    \dot{\bm r} = \tilde{\bm v} + \frac{e}{\hbar} \bm E \times \bm \Omega + \frac{e}{\hbar} (\bm{\dot r} \times  \bm B) \times \bm \Omega=\tilde{\bm v} + \frac{e}{\hbar} \bm E \times \bm \Omega - \frac{e}{\hbar} \bm{\dot r} ({\bm B} \cdot \bm \Omega)+\frac{e}{\hbar} \bm B (\bm{\dot r}\cdot \bm \Omega).
    \label{rdot}
\end{align}
Substituting Eq. (\ref{eq:EOM-1}) in the last term and using $(\dot{\bm k} \times \bm \Omega) \cdot \bm \Omega=0$, we obtain:
\begin{align}
    \bm{\dot r}  = D^{-1}[\tilde{\bm v} + \frac{e}{\hbar} \bm E \times \bm \Omega +  \frac{e}{\hbar} ( \bm \Omega \cdot  \tilde{\bm v}) \bm B]. 
\end{align}
The current density from Eq.~(\ref{Eq:current}) comes out to be
\begin{align}
     j_i = -e \int \frac{d^d \bm k}{(2\pi)^d}  \big[\tilde{v}_i + \frac{e}{\hbar}  (\bm E \times \bm \Omega)_i + \frac{e}{\hbar} (\bm \Omega \cdot \tilde{\bm v}) B_i \big] \big[ f_0(\tilde{\varepsilon}) + \delta f_1 + \delta f_2\big]. 
\end{align}
This expression for the current is valid to $\mathcal{O}(\tau^2)$. 

Assuming weak external fields, we focus on current responses that are (i) linear in the electric field alone and (ii) linear in both electric and magnetic fields, i.e., proportional to $EB$. The former comes out to be
\begin{align}
    j_i (\propto E)= -\frac{e^2}{\hbar} \int \frac{d^d \bm k}{(2\pi)^d}   (\bm E \times \bm \Omega)_i f_0  -\tau \frac{e^2}{\hbar} \int \frac{d^d \bm k}{(2\pi)^d} v_i v_\ell E_\ell \frac{\partial f_0}{\partial \varepsilon}.
    \label{eq.jE}
\end{align}
The first term here represents the anomalous Hall current arising from the Berry curvature. Notably, this contribution is independent of the relaxation time -- it appears even in the absence of scattering processes. The second term is the conventional Drude response, parallel to the applied electric field and linear in $\tau$.

The current proportional to $EB$ is given by
\begin{align}
 j_i (\propto E B)=&   \tau \frac{e^2}{\hbar} \int \frac{d^d \bm k}{(2\pi)^d} \left( \frac{\partial \bm m}{\partial k_i} \cdot \bm B v_\ell E_\ell - v_i E_\ell \frac{\partial {\bm m}}{\partial k_\ell}\cdot \bm B \right) \frac{\partial f_0}{\partial \varepsilon}
 +\tau e^2 \int \frac{d^d \bm k}{(2\pi)^d} v_i E_\ell v_\ell ({\bm m} \cdot \bm B) \frac{\partial^2 f_0}{\partial \varepsilon^2} \nonumber\\
  &  +\tau \frac{e^3}{\hbar} \int \frac{d^d \bm k}{(2\pi)^d} v_i E_\ell v_\ell (\bm B \cdot \bm \Omega) \frac{\partial f_0}{\partial \varepsilon}- \tau \frac{e^3}{\hbar} \int \frac{d^d \bm k}{(2\pi)^d} (\bm \Omega \cdot \bm v) B_i  E_\ell v_\ell   \frac{\partial f_0}{\partial \varepsilon} \nonumber\\
  &-\tau^2 \frac{e^3}{\hbar} \int \frac{d^d \bm k}{(2\pi)^d} v_i E_n (\bm v \times \bm B)_\ell \frac{\partial v_n}{\partial k_\ell} \frac{\partial f_0}{\partial \varepsilon}.\label{eq.jEB}
\end{align}
The first and second integrals involve $\bm m_{\bm k}$, the magnetic moment of the electronic wavepacket. 
The third and fourth integrals involve the Berry curvature, encoding the effect of band topology on transport. They depend on the relative orientation of $\bm B$ with respect to $\bm\Omega$ and $\bm v$. These expressions are in agreement with Ref.~\onlinecite{YangPRB2023}, which discusses longitudinal components of magnetoconductivity. 
The last term is the conventional Hall conductivity, describing a transverse current arising due to the Lorentz force. 

We emphasize that that the above expressions for the current have been derived assuming weak electric and magnetic fields as well as a short relaxation time, which precludes a simple extension of our results to the clean limit. Some of the assumptions can be relaxed when calculating the conventional longitudinal and Hall responses, without the Berry curvature effects. In particular, the conventional Hall conductivity can be calculated for arbitrary relaxation rate and the magnetic field strength, see e.g., Ref. \cite{Abr-book}. 



The conductivity expressions in Eqs.~(7-11) of the main text follow from Eqs.~(\ref{eq.jE}) and (\ref{eq.jEB}) with the assumption that the magnetic field is aligned along the $z$ axis. In addition, we have a contribution to the magnetoconductivity tensor of the form
\begin{align}
    \alpha_{xy}^{m} = \tau \frac{e^2}{\hbar} \int \frac{d^d\bm k}{(2\pi)^d} \left( \bm v \times {\bm \nabla}_{\bm k} m_z \right)_z \frac{\partial f_0}{\partial \varepsilon}+ \tau e^2 \int \frac{d^d\bm k}{(2\pi)^d}\,v_xv_ym_z\,\frac{\partial^2 f_0}{\partial \varepsilon^2}, 
    \label{OMM}
\end{align}
where $m_{z}$ is the magnetic moment of the wavepacket along $z$. 


\section{Berry curvature and magnetic moment under $C_4K$} \label{S3}

We consider a two-dimensional material that breaks $C_4$ and $K$ symmetries, but preserves $C_4K$. This also imbues the material with $C_2$ symmetry, as applying $C_4K$ twice results in $C_2$. The symmetry properties of the velocity $\bm{v}$, $\Omega_z$ and $m_z$ are summarized in Table~\ref{table1}. 

For instance, let us examine the action of $C_4K$ on the Berry curvature. We begin with the semiclassical equations of motion, Eq. (\ref{eq:EOM-1}), in the absence of magnetic moments,
\begin{align}
\dot{x}= v_x + \Omega_{z} \dot{k}_y,\qquad \dot{y}= v_y - \Omega_{z} \dot{k}_x. 
\end{align}
Under $C_4K$, we have 
$\dot{x} \rightarrow -\dot{y}, ~v_x \rightarrow -v_y,~ \dot{k}_y \rightarrow -\dot{k}_x$.
Therefore, in order to preserve the equations of motion, we must have $\Omega_{z} \rightarrow -\Omega_{z}$. 

Similarly, we can find how $C_4K$ acts on the magnetic moment. The semiclassical equations of motion in the presence of nonzero magnetic moments take the form
\begin{align}
\dot{x}= v_x -\frac{\partial m_z}{\partial k_x} B_z+ \Omega_{z} \dot{k}_y,\qquad \dot{y}= v_y -\frac{\partial m_z}{\partial k_y} B_z - \Omega_{z} \dot{k}_x. 
\end{align}
Under $C_4K$, we have $\Omega_{z} \rightarrow -\Omega_{z}$. In order to preserve the equations of motion, we must have $m_z \rightarrow -m_z$.

\begin{table}[]
    \centering
    \begin{tabular}{c|c|c|c|c}
       &$v_x$ & $v_y$  & $\Omega_z$ & $m_z$  \\ 
       \hline $K$ & $-v_x$ & $-v_y$  & $-\Omega_z$ &  $-m_z$  \\
      \hline   $C_2$& $-v_x$ & $-v_y$ & $\Omega_z$&  $m_z$ \\
      \hline   $C_4$& $v_y$ & $-v_x$ & $\Omega_z$& $m_z$ \\
      \hline   $C_4 K$& $-v_y$ & $v_x$& $-\Omega_z$&  $-m_z$ \\
      \hline
    \end{tabular}
    \caption{Symmetry properties of various physical quantities under 
    different symmetry operations.}
    \label{table1}
\end{table}


\section{Tight-binding model} 

To better understand the origin of various terms in the Hamiltonian, we construct a tight-binding model that is consistent with $C_4K$ symmetry. We start with a 2D square lattice, as shown in Fig.~2 in the main text. Apart from the usual hopping processes, we have spin-dependent hopping between next-nearest neighbours. These could arise from the Rashba spin-orbit coupling. We introduce ``altermagnetic order parameters'', $J_1$ and $J_2$. They encode preferential hopping of each spin along nearest and next-nearest neighbour bonds. Crucially, the $J_1$ and $J_2$ processes break $C_4$ and $K$, but preserve $C_4 K$. The Hamiltonian is given by
\begin{align}
    H =& -\frac{t}{2}\sum_{i,\alpha}\big(c^\dagger_{i,\alpha} c_{i+x, \alpha} +c^\dagger_{i,\alpha} c_{i+y, \alpha} + H. c.  \big)  \nonumber\\
    &-i \frac{\lambda}{2} \sum_{i,\alpha \beta}(c^\dagger_{i,\alpha} {\tau}_{2,\alpha\beta} c_{i-x+y,\beta}-c^\dagger_{i-x+y,\beta} {\tau}_{2,\alpha\beta} c_{i,\alpha})  \nonumber\\ 
     &-i \frac{\lambda}{2}\sum_{i,\alpha \beta} (c^\dagger_{i,\alpha} {\tau}_{1,\alpha\beta}  c_{i+x+y,\beta}-c^\dagger_{i+x+y,\beta} {\tau}_{1,\alpha\beta} c_{i,\alpha}) \nonumber\\
    &+ \frac{J_1}{2} \sum_{i,\alpha}\big(c^\dagger_{i,\alpha} {\tau}_{3,\alpha \beta} c_{i+x, \beta} -c^\dagger_{i,\alpha} {\tau}_{3,\alpha \beta} c_{i+y, \beta} + H. c. \big) \nonumber\\ 
&+\frac{J_2}{2} \sum_{i,\alpha}\big(c^\dagger_{i,\alpha} {\tau}_{3,\alpha \beta} c_{i-x+y, \beta} -c^\dagger_{i,\alpha} {\tau}_{3,\alpha \beta} c_{i+x+y, \beta} + H. c.  \big),
\end{align}
where $c^\dagger_{i,\alpha}$ creates a particle with spin $\alpha$ at lattice site $i$, $t$ denotes the hopping parameter, and $\lambda$ corresponds to the spin-orbit coupling. In momentum space, this Hamiltonian takes the form 
\begin{align}
    \hat H(\bm k) =& -t \big(\cos    k_x + \cos    k_y \big) \hat\tau_0 +\frac{\lambda}{2} \left[\sin    \left(k_x+k_y\right)\hat\tau_1 + \sin    \left(k_y-k_x\right)\hat\tau_2\right] \nonumber\\ 
  &+ \left[ J_1(\cos k_x - \cos k_y) + J_2 \sin k_x\sin k_y \right]\hat\tau_3,
  \label{Eq:HamiltonianMomentum} 
\end{align}
where we have set the lattice constant to unity. 


\begin{figure}
\centering
\begin{tabular}{ccc}
\begin{overpic}
[width=0.31\linewidth]{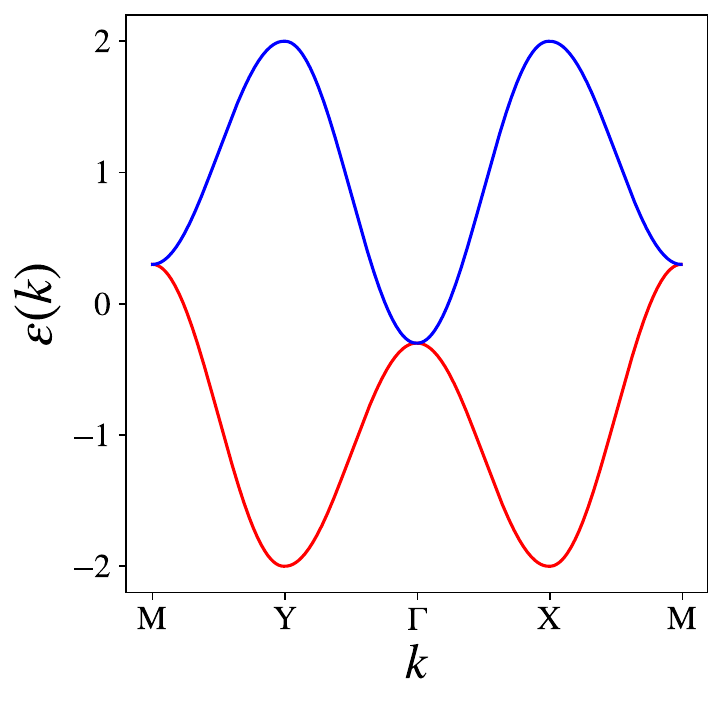}
\put(0,92){\rm{(a)}} 
\end{overpic} & 
\begin{overpic}
[width=0.32\linewidth]{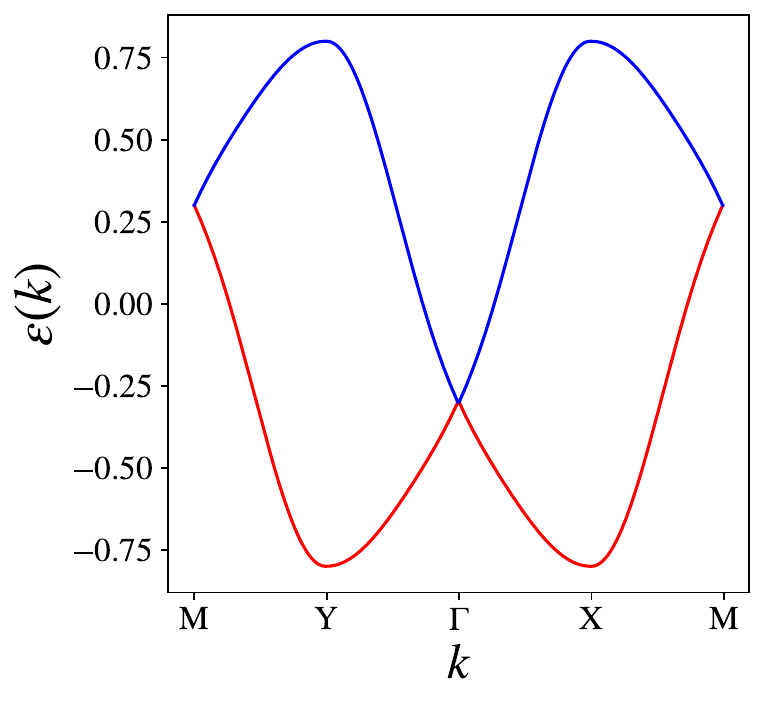}
\put(0,92){\rm{(b)}} 
\end{overpic}
 & 
\begin{overpic}
[width=0.31\linewidth]{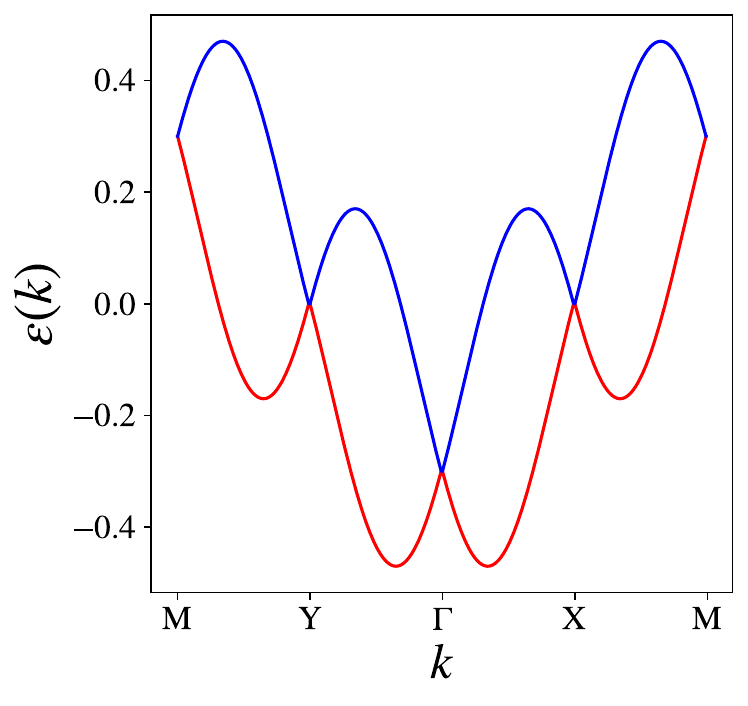}
\put(0,92){\rm{(c)}} 
\end{overpic}
\end{tabular}
\caption{Band structures for different parameter regimes. Panel (a) shows the case $\lambda=0$ with $J_1 = J_2 \neq 0$; panel (b) corresponds to $\lambda = J_1 = J_2$; and panel (c) corresponds to $\lambda \neq 0$ with $J_1 = J_2 = 0$. The parameters used are $t=0.15$, $\lambda=0.4$ and $J_1=J_2=0.4$, all in arbitrary energy units.
}
\label{fig:bands}
\end{figure}

\begin{figure}
\centering
\begin{tabular}{ccc}
\begin{overpic}
[width=0.31\linewidth]{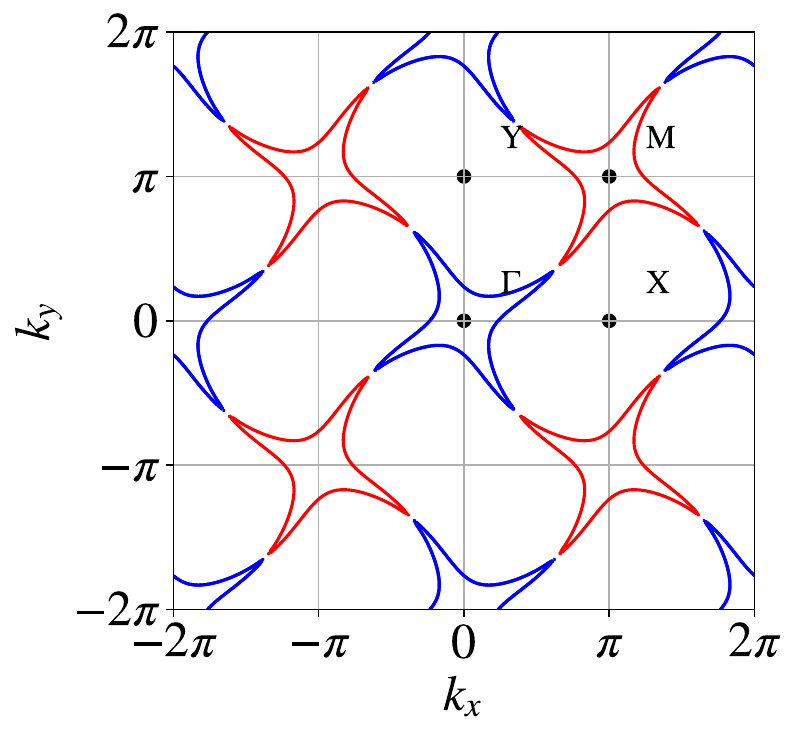}
\put(0,92){\rm{(a)}} 
\end{overpic} & 
\begin{overpic}
[width=0.32\linewidth]{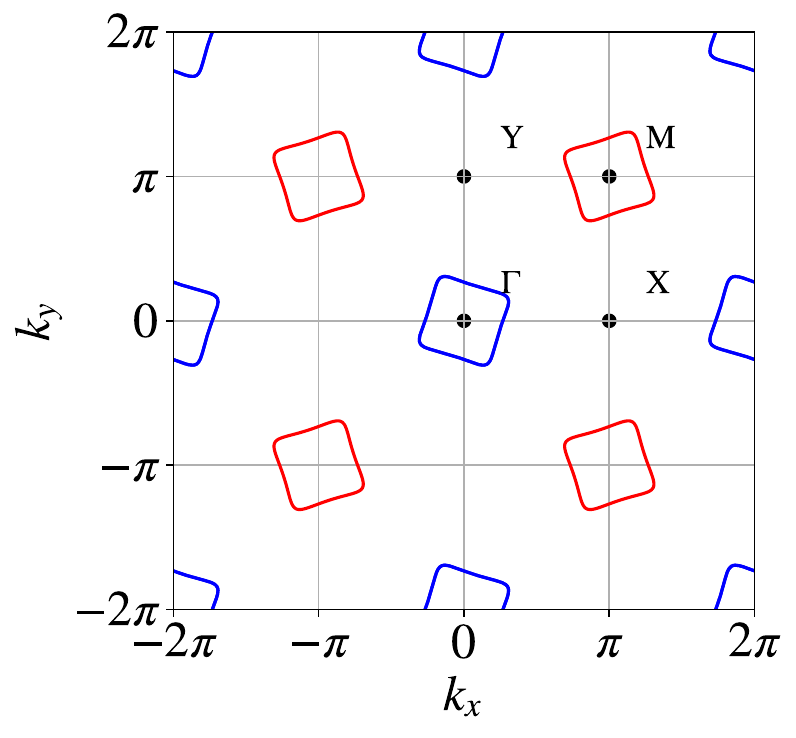}
\put(0,92){\rm{(b)}} 
\end{overpic}
 & 
\begin{overpic}
[width=0.31\linewidth]{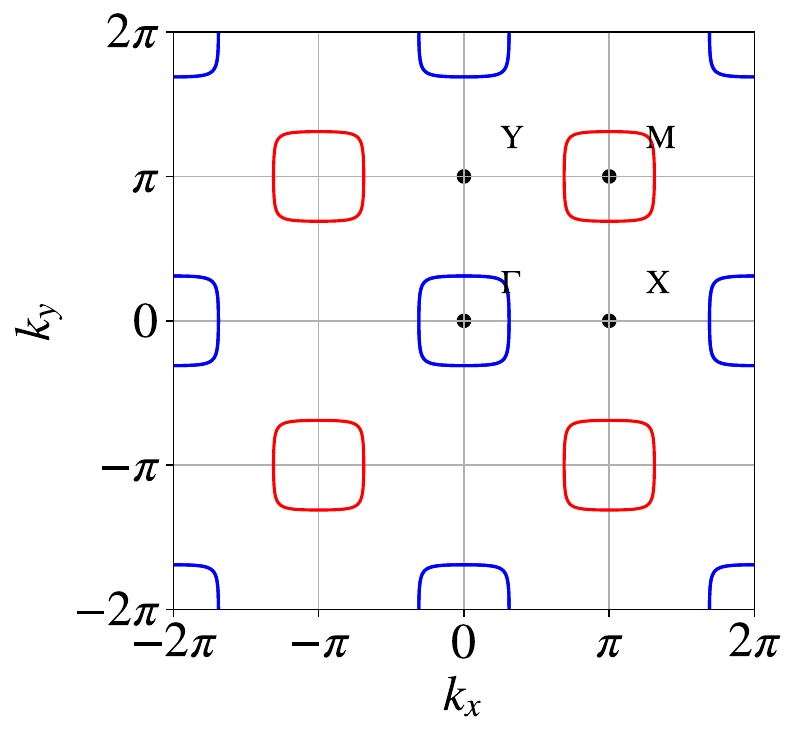}
\put(0,92){\rm{(c)}} 
\end{overpic}
\end{tabular}
\caption{ Fermi surface pockets for various parameter regimes at $\mu=0$. Panel (a) shows the case $\lambda=0$ with $J_1 = J_2 \neq 0$. Panel (b) corresponds to $\lambda = J_1 = J_2$. Finally, panel (c) corresponds to $\lambda \neq 0$ with $J_1 = J_2 = 0$. The parameters used are $t=0.15$, $\lambda=0.4$ and $J_1=J_2=0.4$, all in arbitrary energy units.  }
\label{fig:Fermi}
\end{figure}

Upon diagonalizing this Hamiltonian, the energy eigenvalues are obtained as 
\begin{align}
&\epsilon_\pm({\bm k}) =\delta \pm \frac{1}{2} \sqrt{\xi + \beta+\gamma},
\label{eq.eps}
\end{align}
where
\begin{align}
&\delta = -t (\cos    k_x + \cos    k_y), \nonumber\\
&   \xi = 4 J_1^2 + J_2^2 + \lambda^2 - 8  J_1^2 \cos    k_x \cos    k_y, \nonumber\\ 
&\beta = (2 J_1^2 - J_2^2) (\cos2     k_x + \cos 2    k_y) + (J_2^2 - \lambda^2) \cos2    k_x \cos2    k_y, \nonumber\\ 
&\gamma =  4 J_1 J_2 (\sin2    k_x \sin    k_y -  \sin    k_x \sin 2    k_y).\nonumber
\end{align}
Here $\epsilon_\pm({\bm k})$ denote the energy of upper and lower bands. Fig.~\ref{fig:bands} plots the energy bands along the $M-Y-\Gamma-X-M$ contour in the Brillouin zone. The two high-symmetry points of Eq.~(\ref{Eq:HamiltonianMomentum}), $\Gamma=(0,0)$ and $M=(\pi,\pi)$ host gapless Dirac nodes with $\epsilon_\Gamma =-2t$ and $\epsilon_M = 2t$. When $t$ is small and the chemical potential is close to zero, we obtain two Fermi pockets around each Dirac point, see Fig.~\ref{fig:Fermi}. 

To connect with the long-wavelength form, we examine the form of the tight-binding Hamiltonian near the Dirac nodes. Near $\Gamma$, keeping terms up to second order in ${\bm k}$, the Hamiltonian (\ref{Eq:HamiltonianMomentum}) takes the form 
\begin{align}
    H_{\Gamma}({\bm k}) =& -\frac{t  }{2}(4- k^2) \hat{\tau}_0 + \frac{\lambda  }{2} (k_x+k_y) \hat{\tau}_1 + \frac{\lambda    }{2} (k_y-k_x) \hat{\tau}_2 -\left[\frac{J_1   }{2} (k^2_x-k^2_y) - J_2    k_x k_y \right] \hat{\tau}_3, 
\label{SupplEq:H-Gamma}
\end{align}
whereas near the $M$ point we have
\begin{align} 
    H_M(\bm k) = \frac{t   }{2}(4- k^2)\hat{\tau}_0 + \frac{\lambda    }{2} (k_x+k_y) \hat{\tau}_1 +\frac{\lambda   }{2}(k_y-k_x)\hat{\tau}_2+ \left[\frac{J_1   }{2}(k_x^2-k_y^2) + J_2    k_xk_y \right] \hat{\tau}_3. 
\label{SupplEq:H-M}
\end{align} 
Near the two Dirac points, the band dispersions of Eq.~(\ref{eq.eps}) take the form 
\begin{align}
    \epsilon_{\pm}^\Gamma(\bm k) =  -\frac{t}{2}(4 -k^2) \pm \frac{1}{2} \sqrt{ \big[J_1 ( k_x^2 - k_y^2 ) - 2J_2 k_x k_y\big]^2 + 2\lambda^2 k^2}
\label{SupplEq:epsilon-pm-Gamma}
\end{align}
and 
\begin{align}
    \epsilon_{\pm}^{\rm M}(\bm k) =  \frac{t}{2}(4 -k^2) \pm \frac{1}{2} \sqrt{ \big[J_1 ( k_x^2 - k_y^2 ) + 2 J_2 k_x k_y\big]^2 + 2\lambda^2 k^2}.
\label{SupplEq:epsilon-pm-M}
\end{align}

\section{Berry curvature and magnetic moment in the two-band model}

In a two-level system, the Hamiltonian can be represented as $\hat H = d_0({\bm k})\hat\tau_0 + {\bm d}(\bm k)\cdot\hat{\bm\tau}$. Here, we take the Pauli matrices to act on the physical spin of the electron. The associated Berry curvature, orbital and spin magnetic moments are given by~\cite{BerryReview} 
\begin{align}
    \Omega_{z,\pm} = \pm \epsilon_{ij z }\frac{ 1}{2|{\bm d}(\bm k)|^3} {\bm d}(\bm k) \cdot \left[\frac{\partial{\bm d}(\bm k)}{\partial k_i} \times \frac{\partial{\bm d}(\bm k)}{\partial k_j} \right],
\label{Eq:BerryC}
\end{align}
\begin{align}
    m_{z,\pm}^{\rm orb} =\mp \epsilon_{ij z }\frac{ e}{4 \hbar}\frac{1}{ |{\bm d}(\bm k)|^2} {\bm d}(\bm k) \cdot \left[\frac{\partial{\bm d}(\bm k)}{\partial k_i} \times \frac{\partial{\bm d}(\bm k)}{\partial k_j} \right],
\label{Eq:m-orb-z}
\end{align}
and
\begin{align}
    {\bm m}^{\rm s}_\pm = \pm \frac{1}{2} \mu_{\rm B} g \frac{\bm d(\bm k)}{|\bm d(\bm k)|},  
\end{align}
where $\mu_{\rm B}$ is the Bohr magneton, $g$ is the Land\'e $g$-factor, and $\pm$ indicates the upper and lower bands. 

\begin{figure}
\centering
\begin{tabular}{ccc}
\begin{overpic}
[width=0.31\linewidth]{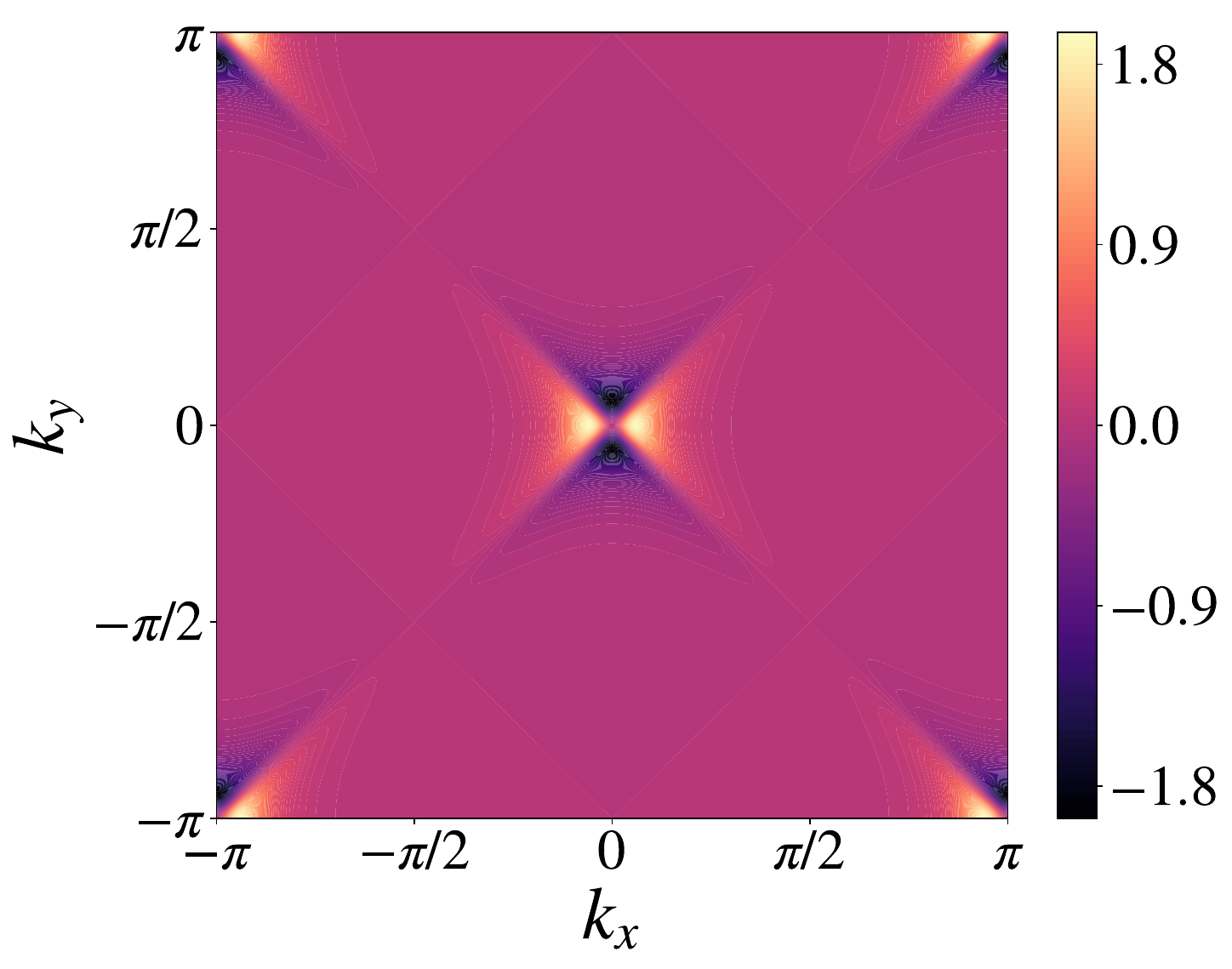}
\put(0,84){\rm{(a)}} 
\end{overpic} & 
\begin{overpic}
[width=0.31\linewidth]{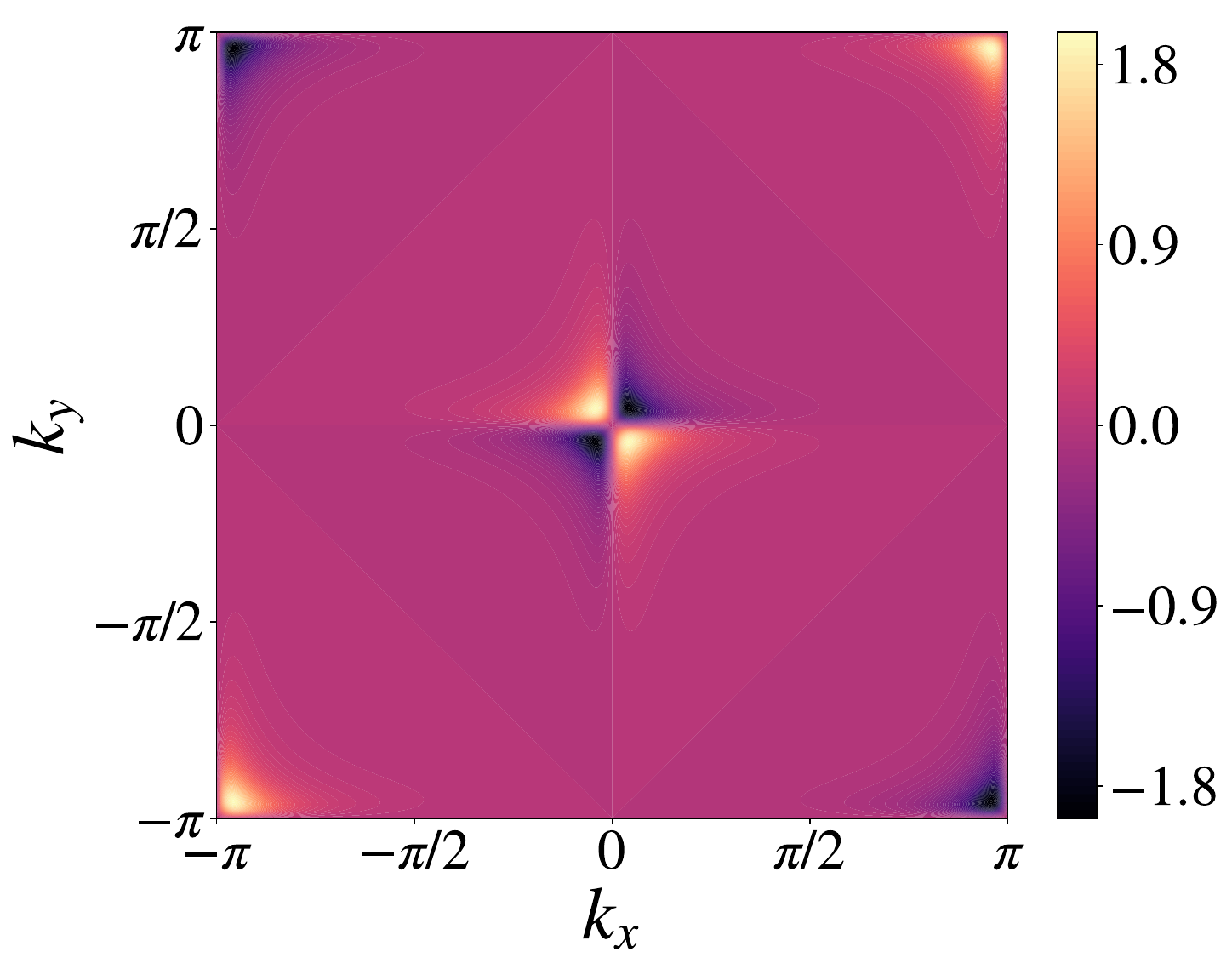}
\put(0,84){\rm{(b)}} 
\end{overpic}
 & 
\begin{overpic}
[width=0.31\linewidth]{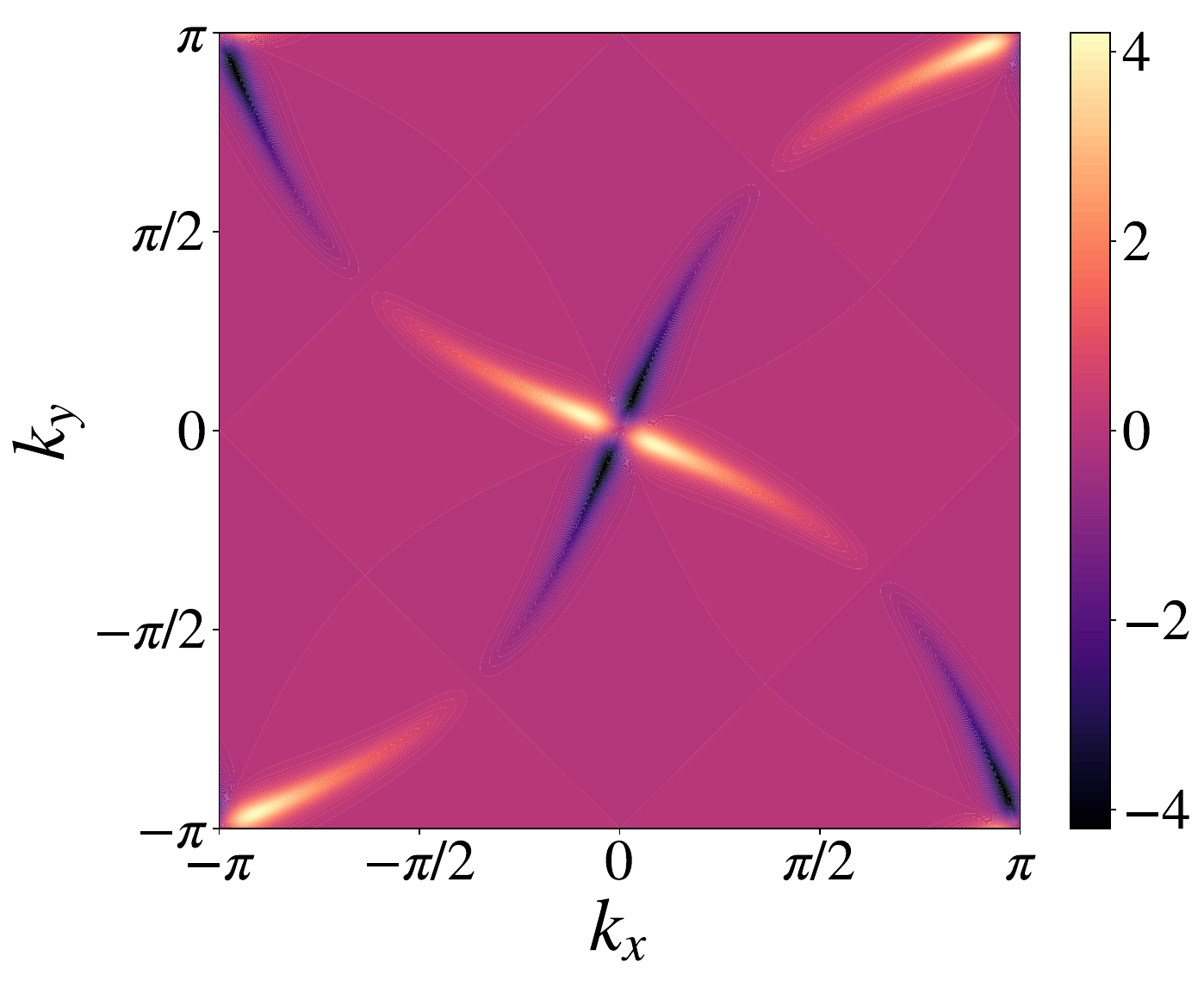}
\put(0,84){\rm{(c)}} 
\end{overpic}
\end{tabular}
\caption{Quadrupole-like distribution of the Berry curvature in Brillouin zone for the upper band. (a)  $J_1 \neq 0$ and $J_2=0$, (b) $J_1=0$ and $J_2 \neq 0 $ and (c) $J_1 = J_2 \neq 0 $. The $C_4K$ symmetry is clearly seen. The distributions in (a) and (b) show $d_{x^2-y^2}$ and $d_{xy}$ character, respectively.
}
\label{fig:Berryquad}
\end{figure}

\begin{figure}
\centering
\begin{tabular}{ccc}
\begin{overpic}
[width=0.31\linewidth]{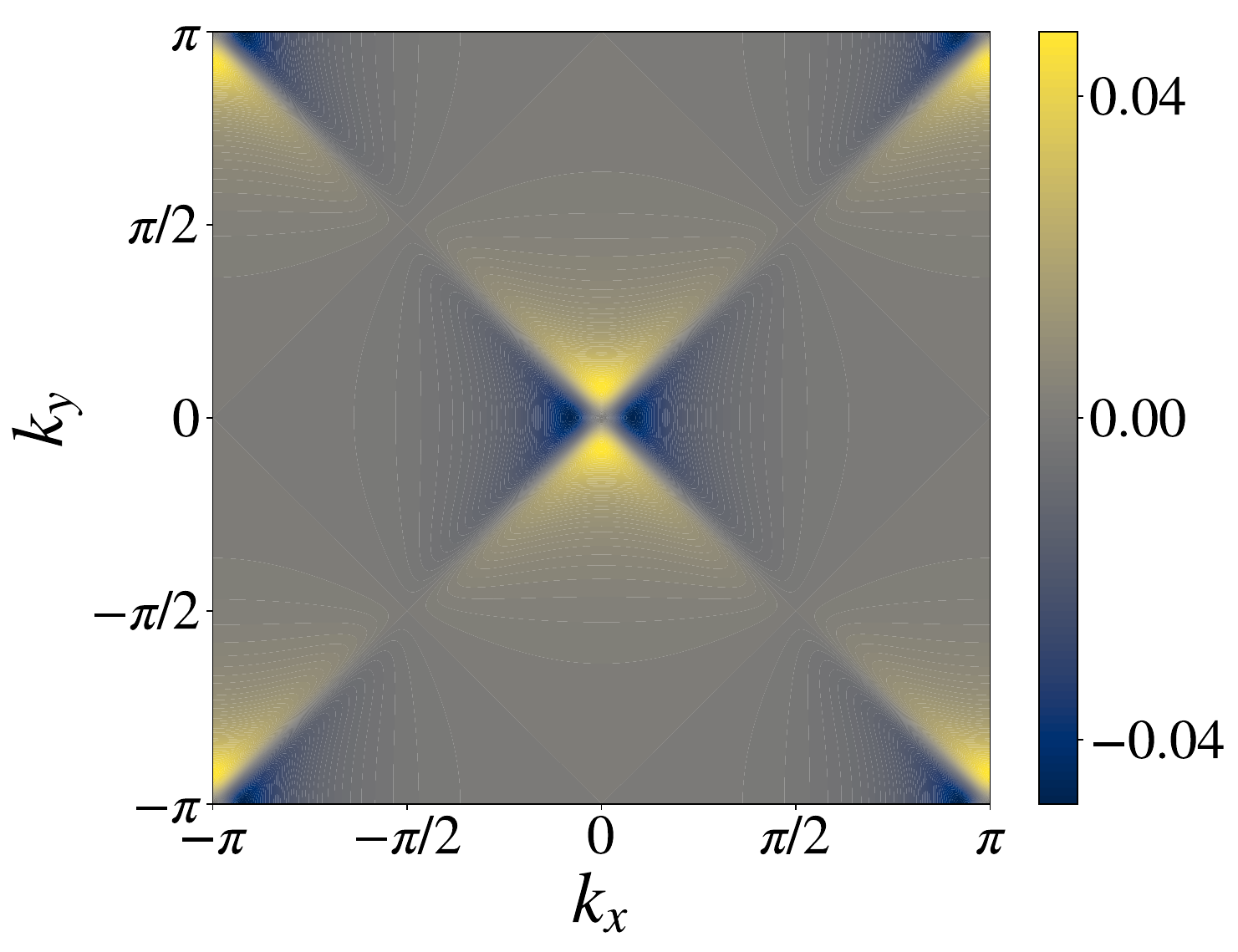}
\put(0,82){\rm{(a)}} 
\end{overpic} & 
\begin{overpic}
[width=0.31\linewidth]{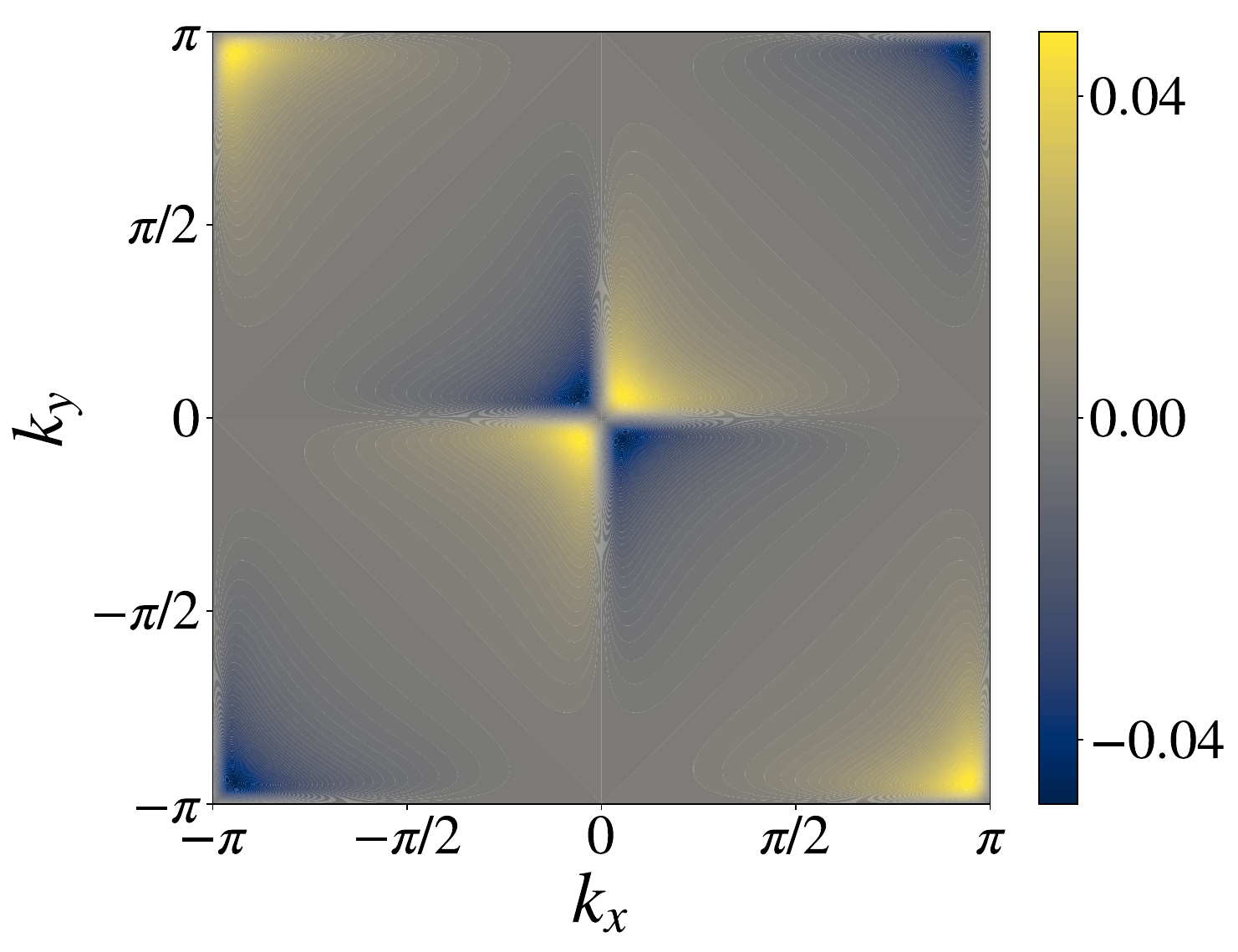}
\put(0,82){\rm{(b)}} 
\end{overpic}
 & 
\begin{overpic}
[width=0.31\linewidth]{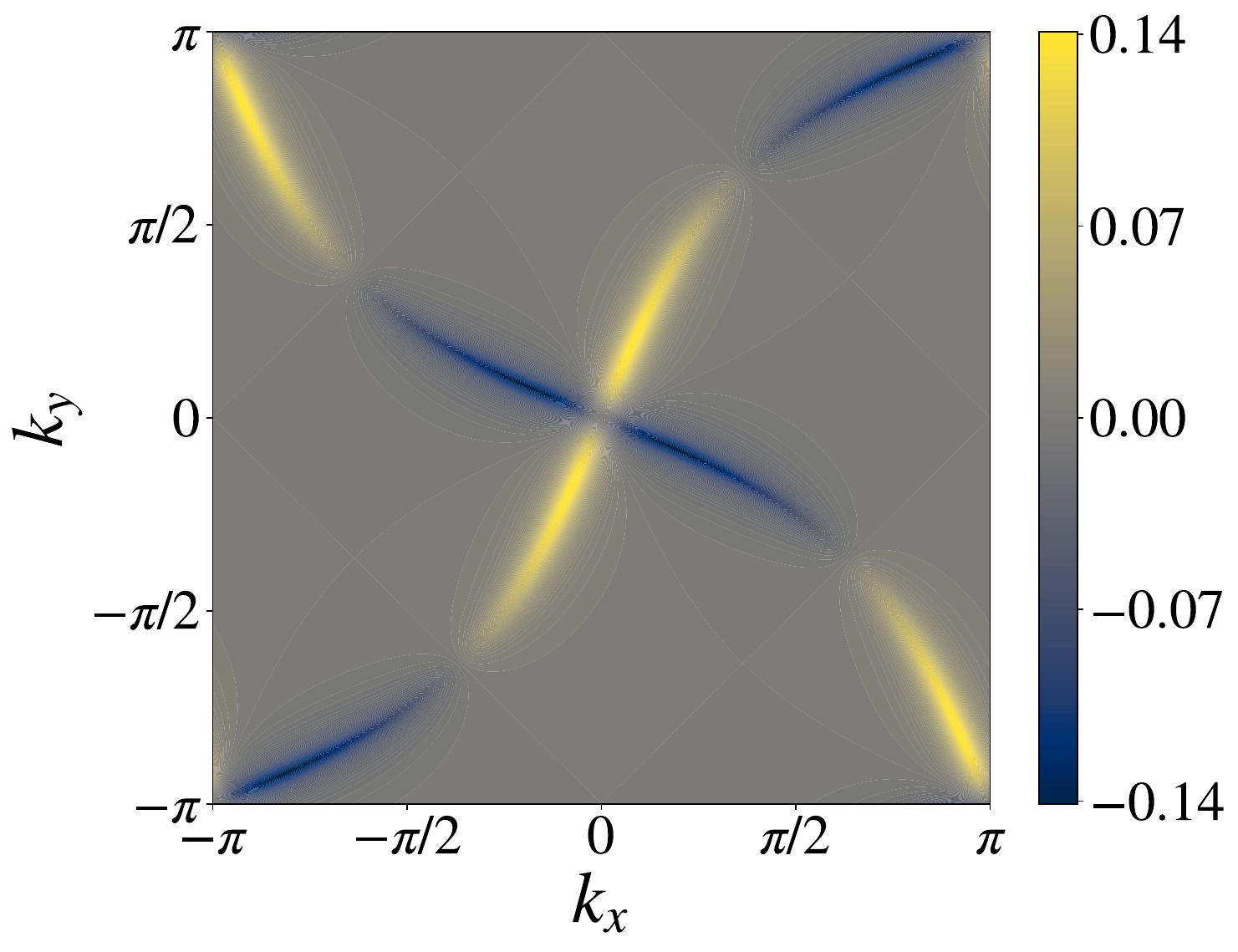}
\put(0,82){\rm{(c)}} 
\end{overpic}
\end{tabular}
\caption{Quadrupole-like distribution of the orbital magnetic moment in Brillouin zone for the upper band. (a)  $J_1 \neq 0$ and $J_2=0$, (b) $J_1=0$ and $J_2 \neq 0 $ and (c) $J_1 = J_2 \neq 0 $. 
}
\label{fig:Magnetquad}
\end{figure}

In the tight-binding model described above, the Berry curvature at an arbitrary point in the Brillouin zone comes out to be
\begin{align}
    \Omega_{z,\pm} = \mp\frac{\lambda^2  (\cos    k_x + \cos    k_y) [ J_1  (\cos    k_x- \cos    k_y) + J_2  \sin    k_x \sin    k_y ]}{8\left(\lambda^2 \big(1- \cos    k_x \cos    k_y \big) + [J_1  (\cos    k_x - \cos    k_y)  + J_2 \sin    k_x \sin    k_y]^2\right)^{3/2}}. 
    \label{Eq:BerryLattice+}
\end{align}
The orbital and spin magnetic moments are given by 
\begin{align}
   m_{z,\pm}^{\rm orb}  = \pm \frac{e}{16\hbar }  \frac{ \lambda^2  (\cos    k_x + \cos    k_y) [ J_1 (\cos    k_x- \cos    k_y) + J_2  \sin    k_x \sin    k_y ] }{\lambda^2 \big(1- \cos    k_x \cos    k_y \big)  +  [ J_1 (\cos    k_x - \cos    k_y)  + J_2 \sin    k_x \sin    k_y]^2 }
\end{align} 
and
\begin{align}
    m^{\rm s}_{z,\pm} = \pm \frac{\mu_{\rm B} g}{2}  \frac{ J_1(\cos    k_x - \cos    k_y) + J_2 \sin    k_x\sin    k_y }{\left(\lambda^2 \big(1- \cos    k_x \cos    k_y \big)  +  [ J_1 (\cos    k_x - \cos    k_y)  + J_2 \sin    k_x \sin    k_y]^2 \right)^{1/2}}
\end{align}
We immediately see that under $C_4K$, $\Omega_{z,\pm} \to -\Omega_{z,\pm} $ and $m_{z,\pm} \to -m_{z,\pm} $, as previously argued on symmetry grounds, see Sec.~\ref{S3}. Additionally we find that $\Omega_{z,+} = - \Omega_{z,-}$ and $m_{z,+} = - m_{z,-}$. 
Figs.~\ref{fig:Berryquad} and \ref{fig:Magnetquad} show $ \Omega_{z,+}$ and $m_{z,\pm}^{\rm orb}$ over the Brillouin zone for three representative values of $J_1$ and $J_2$. The plots exhibit clear quadrupole-like distributions, indicative of $C_4K$ symmetry. 

In the immediate vicinities of $\Gamma$ and $M$ points, the Berry curvature comes out to be
\begin{align}
   \Omega^{\Gamma}_{z,\pm} =  \pm \frac{[J_1  (k_x^2 - k_y^2) - 2 J_2    k_x k_y]  \lambda^2   }{\left( [J_1  (k_x^2- k_y^2)-2J_2    k_xk_y]^2  + 2 \lambda^2  k^2 \right)^{3/2}}
\end{align}
and
\begin{align}
   \Omega^{M}_{z,\pm} =  \mp \frac{[J_1    (k_x^2 - k_y^2)  + 2 J_2    k_x k_y]  \lambda^2   }{\left( [J_1  (k_x^2 - k_y^2)+2J_2   k_xk_y]^2  + 2\lambda^2  k^2 \right)^{3/2}}.
\end{align}
For the orbital magnetic moment around  $\Gamma$ and $M$ we obtain:
\begin{align}
   m^{{\rm orb},\Gamma}_{z,\pm} =  \mp \frac{e}{2\hbar} \frac{  \lambda^2 [  J_1(k_x^2 - k_y^2) - 2 J_2   k_x k_y]     }{  [ J_1 (k_x^2 - k_y^2)-2 J_2  k_xk_y]^2  + 2 \lambda^2  k^2},
\end{align}
and
\begin{align}
   m^{{\rm orb},M}_{z,\pm} =  \pm \frac{e}{2\hbar} \frac{  \lambda^2 [   J_1 (k_x^2 - k_y^2)  + 2 J_2   k_x k_y]      }{  [  J_1(k_x^2 - k_y^2)+2  J_2 k_xk_y]^2  + 2\lambda^2  k^2 }.
\end{align}
Finally, the spin magnetic moment around $\Gamma$ and $M$ is given by
\begin{align}
     m^{{\rm s},\Gamma}_{z,\pm} = \mp \frac{\mu_{\rm B} g}{2}   \frac{J_1  (k^2_x-k^2_y)-2J_2 k_x k_y }{\left(  [ J_1 (k_x^2 - k_y^2)-2 J_2  k_xk_y]^2  + 2 \lambda^2  k^2 \right)^{1/2}},
\end{align}
and
\begin{align}
     m^{{\rm s},M}_{z,\pm} = \pm \frac{\mu_{\rm B} g}{2}   \frac{J_1  (k^2_x-k^2_y) +2J_2  k_x k_y }{\left(  [  J_1(k_x^2 - k_y^2)+2  J_2 k_xk_y]^2  + 2\lambda^2  k^2 \right)^{1/2}}. 
\end{align}

\section{Conductivity integrals}

We have Dirac points at $\Gamma$ and $M$, reflecting a Kramers-like degeneracy enforced by $C_4K$ symmetry. When the Fermi energy is close to these Dirac points, two small Fermi pockets emerge. In the vicinity of the Dirac points, we express the momentum in polar coordinates as $(k_x, k_y) = k(\cos\varphi, \sin\varphi)$. 

The band dispersion of the lower band near the $\Gamma$ point, as detailed in Eqs. (\ref{SupplEq:epsilon-pm-Gamma}) and (\ref{SupplEq:epsilon-pm-M}), is given by
\begin{align}
    \epsilon (\bm k) = -2t + \frac{tk^2}{2} 
    - \frac{\sqrt{2}}{2}\lambda k \bigg[1 + \frac{1}{2} \frac{\tilde J^2(\varphi)k^2}{2\lambda^2}  + \dots \bigg],
\end{align}
where $\tilde J(\varphi) = J_1\cos(2\varphi) - J_2\sin(2\varphi)$. The leading term in this expression describes an isotropic (circular) Fermi surface. We have anisotropies that are proportional to $J_{1,2}^2/\lambda^2$.  Below, we calculate current contributions assuming that $J_1,J_2\ll\lambda$, which corresponds to nearly-isotropic Fermi surfaces. In order to evaluate currents to the leading order in $J_{1,2}/\lambda$, we need 
\begin{align}
    \Omega_z = -  \frac{\tilde J(\varphi)}{2 \sqrt{2} \lambda   k} \bigg[1-\frac{3}{2}  \frac{\tilde J^2(\varphi)k^2}{2\lambda^2 }  + \dots\bigg], 
\end{align} 
and 
\begin{align}
    m^{\rm orb}_z = -\frac{e}{\hbar} \frac{\tilde J(\varphi) }{4} \bigg[1-  \frac{\tilde J^2(\varphi)k^2}{2\lambda^2 }+ \dots \bigg]. 
\end{align}
and 
\begin{align}
    m^{\rm s}_z = - \mu_{\rm B} g  \frac{\tilde J(\varphi)  k}{2\sqrt{2} \lambda} \bigg[1-\frac{1}{2}  \frac{\tilde J^2(\varphi)k^2}{2\lambda^2 }+ \dots \bigg]. 
\end{align}

Elements of the conductivity tensor involve several integrals over the Brillouin zone, see Eqs. (\ref{eq.jE}), (\ref{eq.jEB}), and (\ref{OMM}). We evaluate these integrals assuming that temperature is much smaller than all other energy scales, such as the Fermi energy, bandwidth, etc. In this limit, the equilibrium distribution function $f_0$ can be viewed as a step-function in energy. After integration by parts, we may write $\partial_\epsilon f_0=-\delta[\epsilon(\bm k)-\mu]  $, where $\mu$ is the chemical potential. Along these lines, each integral in the expressions can be written in the following form:
\begin{equation}
    I = \frac{1}{(2\pi)^2} \int_0^\infty k \, dk \int_0^{2\pi} d\varphi \, F(k, \varphi) \left(\frac{1}{|\partial_k \epsilon(k)|}\right)_{k=k_F} \delta(k-k_F),
\label{suppl-eq:I-theta k}
\end{equation}
where $F(k,\varphi)$ is some function in 2D polar coordinates,
\begin{align}
    |\partial_k \epsilon(k)&|_{k=k_F(\varphi)}=t k_F(\varphi)  - \frac{\sqrt{2}}{2}\lambda - \frac{ 3\tilde J^2(\varphi) k_F^2(\varphi)}{4 \sqrt{2} \lambda}
\end{align}
and the Fermi wave vector is given by
\begin{align}
    k_F(\varphi) \simeq  (\mu + 2 t) \biggl[ -\frac{\sqrt{2}}{\lambda} + \frac{1}{v_0}  \frac{\tilde J^2(\varphi)}{2\lambda^2  }\left(\frac{\mu+2t}{\lambda}\right)^2  \biggr].
\end{align}
Here $v_0= {d \epsilon_0} /dk$ is the isotropic Fermi velocity and $\epsilon_0 =-2t + {tk^2}/{2}
    -\lambda  k/\sqrt{2}$. 
Evaluating the integral over $k$, Eq.~(\ref{suppl-eq:I-theta k}) takes the form 
\begin{align}
   I = \frac{1}{(2\pi)^2} \int_0^{2\pi} d\varphi~ k_F~F(k_F, \varphi) \left(\frac{1}{|\partial_k \epsilon(k)|}\right)_{k=k_F}.
\end{align}

We now evaluate each contribution to the magnetoconductivity tensor. For brevity, we provide explicit expressions only for the $\Gamma$ pocket, to leading order in $J = J_{1,2}/\lambda$. 
To study dependence on temperature, we assume that altermagnetic order sets in via a second-order phase transition at a critical temperature $T_c$. Following standard Landau theory, we suppose that the altermagnetic order parameter is given by
\begin{align}
J(T) = J_0 \sqrt{1 - \frac{T}{T_c}} \quad \text{for } T < T_c,
\label{eq.JT}
\end{align}
where $J_0$ is a constant that depends on material parameters. 
As $T$ approaches $T_c$ from below, the system transitions to a disordered state (with $C_4$ and $K$ symmetries) where altermagnetic order parameter vanishes. We have 
 \begin{align}
    \alpha_{xy}^{\text{D} ,\Gamma} = - \frac{\tau^2e^3t^2 \epsilon_\Gamma^2}{2\pi \hbar^2 \lambda^2}  \bigg[1   - \frac{\sqrt{2}}{2}   \frac{J^2(T)}{\lambda^2} \bigg],
\end{align}

\begin{equation}
 \alpha_{xy}^{\text{B},\Gamma} = \frac{\tau e^3 }{32 \pi \hbar \lambda^2}
\frac{(\lambda^2-2t\epsilon_\Gamma)^2}{\lambda^2+2t\epsilon_\Gamma} J(T),
\end{equation}
and
\begin{align}
\alpha_{xy}^{m,\Gamma}  = \frac{\tau e^3 }{64 \pi \hbar}   \frac{\lambda^2 + 8 t \epsilon_\Gamma}{\lambda^2 +2 t \epsilon_\Gamma} J(T) + \frac{3 \tau e^2 }{32 \pi} g \mu_{\rm B} \frac{\epsilon_\Gamma }{\lambda^2} \frac{\lambda^2 + 4t  \epsilon_\Gamma}{\lambda^2-2 t \epsilon_\Gamma }  J(T) 
\end{align}
where $\epsilon_\Gamma = |2t+\mu|$ is the  energy of the degenerate band at the $\Gamma$ point. Contributions from the vicinity of the M point can be calculated on similar lines. 
We find that the magnitude of $\alpha_{xy}^{m}$ is much smaller than $\alpha_{xy}^{\text{D}}$ and $\alpha_{xy}^{\text{B}}$ for plausible parameter values. 

\begin{figure}
    \centering
    \includegraphics[width=0.48\linewidth]{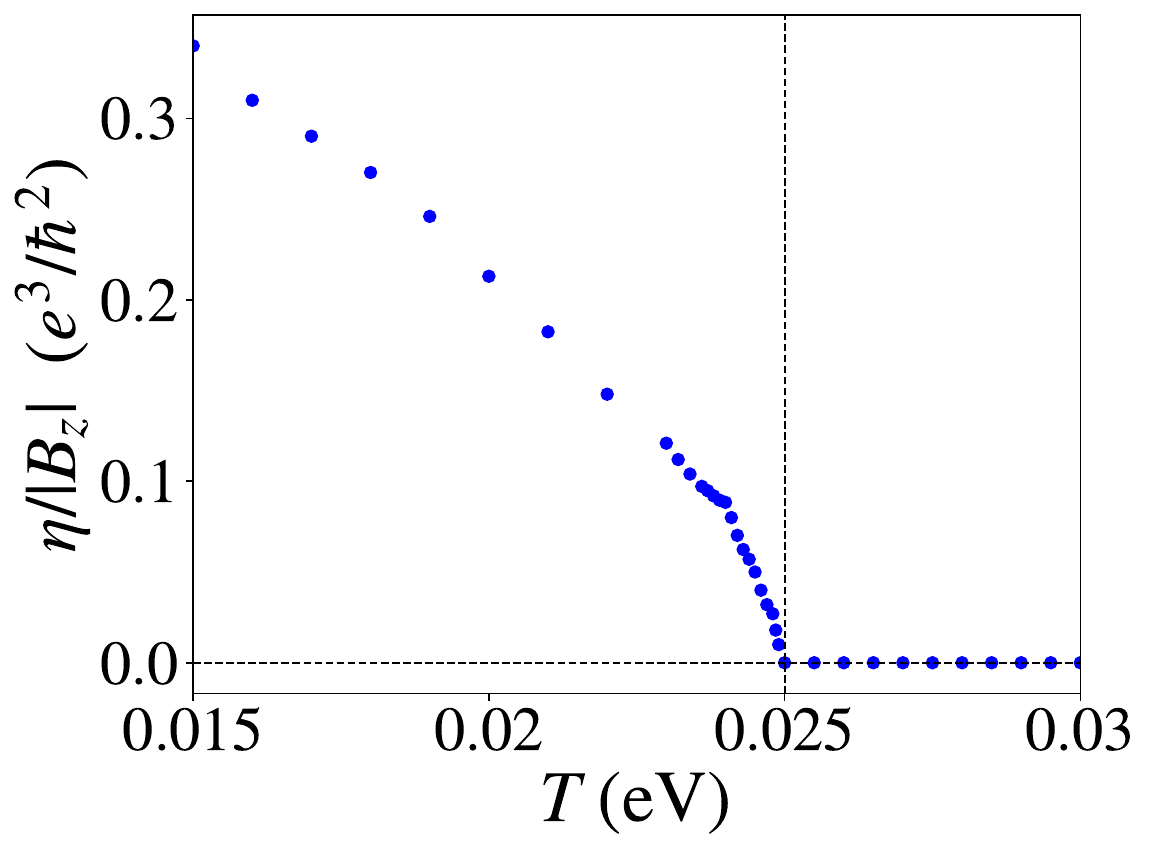}
    \caption{Evolution of the magnetoconductivity anisotropy with temperature. We plot the result calculated directly from the tight-binding model (\ref{Eq:HamiltonianMomentum}). The chemical potential is set at $\mu=0$. The parameters used are the same as the Fig.~4 in the main text, except $J_1=2J_2=J_0=1~eV$. The dashed line indicates the critical temperature $T_c$.}
    \label{fig:SigmaSup}
\end{figure}
Fig.~4 in the main text shows the numerically-calculated magnetoconductivity anisotropy for $J_1=J_2$. In Fig.~\ref{fig:SigmaSup} here, we plot the same quantity for $J_1\neq J_2$. We find the same qualitative result with a square-root singularity at $T_c$.


\clearpage

\end{document}